\documentclass[11pt]{article}

\usepackage{epsf}
\usepackage{dsfont}  
\usepackage{hyperref}

\usepackage{epsfig}
\usepackage{graphicx}
\usepackage{amsfonts}
\usepackage{color}
\usepackage{epstopdf}
\usepackage{graphicx}
\usepackage{amsmath}
\usepackage{amssymb}
\usepackage{amsthm}
\usepackage{natbib}
\usepackage{fancybox}
\usepackage{ulem}
\usepackage{subfigure}
\usepackage{multirow}
\usepackage{algorithm}
\usepackage{algorithmic}
\usepackage{amsmath}
\numberwithin{equation}{section}
\usepackage{enumerate}

\newtheorem{remark}{Remark}[section]
\newtheorem{exam}{\textsc{Example}}[section]

\newtheorem{lemma}{\hspace{6mm}Lemma}[section]

\def\be{\begin{equation}}
\def\ee{\end{equation}}

\newcommand{\pf}{\noindent {\bf Proof} \hspace{2mm}}
\newcommand{\eproof}{$\quad \Box$}

\begin{document}

\title{Delta family approach for the stochastic control problems of utility maximization}

\vskip 4mm

\author{Jingtang Ma\thanks{School of Mathematics and Fintech Innovation Center, Southwestern University of
Finance and Economics, Chengdu, 611130, China (Email: mjt@swufe.edu.cn). The work
was supported by National Natural Science Foundation of
China (Grant No. 12071373) and the Fundamental Research Funds for the Central Universities China (JBK1805001).}, Zhengyang Lu\thanks{School of Mathematics, Southwestern University of Finance and Economics, Chengdu, 611130, China (Email: luzy@smail.swufe.edu.cn).}~ and Zhenyu Cui\thanks{Corresponding author. School of Business, Stevens Institute of Technology, Hoboken,  New Jersey 07030, United States. (Email: zcui6@stevens.edu).}
}

\date{\today}

\maketitle

\begin{abstract}
In this paper, we propose a new approach for stochastic control problems arising from utility maximization. The main idea is to directly start from the dynamical programming equation  and compute the conditional expectation using  a novel representation of the conditional density function through the Dirac Delta function and the corresponding  series representation. We obtain an explicit series representation of the value function, whose coefficients are expressed through integration of the value function at a later time point against a chosen basis function. Thus we are able to set up a recursive integration time-stepping scheme to compute the optimal value function given the known terminal condition, e.g. utility function. Due to tensor decomposition property of the Dirac Delta function in high dimensions, it is straightforward to extend our approach to solving high-dimensional stochastic control problems. The backward recursive nature of the method also allows for solving stochastic control and stopping problems, i.e. mixed control problems.  We illustrate the method through solving some  two-dimensional stochastic control (and stopping) problems, including the case under the classical and rough Heston stochastic volatility models, and stochastic local volatility models such as the stochastic alpha beta rho (SABR) model.
\end{abstract}

\vspace{1.0cm}
\noindent {\bf Keywords:} {stochastic control, Dirac Delta function, Delta sequence, HJB equation}

\medskip

\noindent {\bf JEL classification:} G12, G13, G14, C58

\section{Introduction}
Stochastic control generally refers to the study of dynamical systems subject to random perturbations and depending on the unknown control, and seeks to find the optimal control that maximizes or minimizes certain performance functions. It has  applications in many areas in science, including applied mathematics, engineering, economics, and finance. In finance, it is usually associated with the maximization of expected utility function evaluated at the terminal wealth. There is a vast  literature on expected utility maximization or optimal portfolio choice starting from the seminal work of \cite{Merton1969}. There are two fundamental approaches to solving stochastic control problems: first is the Bellman's dynamical programming principle (DPP), which leads to the Hamilton-Jacobi-Bellman (HJB) partial differential equation (PDE) to characterize the optimal value function; second is the Pontryagin's maximum principle. In recent years there have been considerable developments within both approaches, in particular inspired by problems arising from mathematical finance. The introduction of viscosity solution allows for going beyond the classical verification Bellman approach, and sets a rigorous and suitable framework for studying stochastic control using the DPP approach. On the other hand, the Pontryagin's maximum principle finds a suitable presentation with the backward stochastic differential equations (BSDEs). The viscosity solution and BSDE are related through the PDE representation.

There are various methods for solving stochastic control problems arising from utility maximizations in the literature. One approach is of analytical nature and  focuses on numerically solving the related HJB equation, which is a nonlinear PDE, see \cite{Forsyth2007,White2011}. An alternative approach is of probabilistic nature and utilizes the martingale duality method, see \cite{Karatzas1991,Civitanic1992}.  A recent account of the current literature is provided in \cite{Pham2009} and references therein. An extension to the finite-maturity utility maximization problem is to allow the investor to ``exit" before the maturity in order to achieve the overall maximization of the expected utility. This leads to a mixed stochastic control problem which involves both optimal control and optimal stopping, see \cite{Ka2000}. Although there is extensive literature on the study of one-dimensional stochastic control problems, in contrast, the literature on high-dimensional stochastic control is relatively thin, see \cite{Belo2010} for a Monte Carlo regression approach, and \cite{Han2016}, \cite{Pham2021} for deep learning approach. 
Note that there are also further extensions of the utility maximization problem to allow for (fixed or proportional) transaction costs \citep{Gua2013}  or taxes \citep{Tahar2010}. We focus on the high-dimensional finite-maturity optimal control and mixed control problems in this paper, and leave the consideration of these further extensions to future research.

The motivation behind our proposed approach stems from the following observation: the dynamical programming equation involves the computation of the associated conditional expectation whose payoff function is the optimal value function solved from the previous step, and in the first step the payoff function will be the given utility function.  The traditional approach to derive the HJB equation involves rewriting this conditional expectation as the solution to a corresponding parabolic PDE through the Feynman Kac theorem, whose coefficients depend on the unknown control. Then we ``maximize" the PDE with respect to the unknown control, and substitute the solution back to the PDE to arrive at the nonlinear HJB PDE. We take an alternative route and express the conditional expectation using the conditional density function, for which we find a representation using property of the Dirac Delta function and its series representation.

The contributions of the paper are three-fold. First, we propose a novel time-stepping scheme to solve HJB equations involving recursive integrations at each step. The method is straightforwardly extended to the high-dimensional stochastic control problems. Second, the method can be applied to mixed control problems involving optimal stopping. Third, we demonstrate the accurateness and efficiency of the method in two-dimensional stochastic control problems arising from applications, involving Heston, rough Heston and representative  stochastic local volatility models, e.g. the stochastic alpha beta rho (SABR) model.

The remainder of the paper is organized as follows: Section \ref{sec2} introduces the delta family method for both one-dimensional and high-dimensional stochastic control problems. Section \ref{sec3} illustrates the applications of the method to solving the optimal investment problem under the Heston model, the optimal reinsurance and investment problem under the Heston and rough Heston model, and also a mixed stochastic control problem under general stochastic local volatility models. Section \ref{sec4} concludes the paper.

\section{Delta family approach for the stochastic control problems}\label{sec2}

\subsection{Delta family representation of density function}

In this section, all expectations and distributions are taken under the
physical or real world measure $P$. Consider a one-dimensional Markov process $
\{X_{t}\}_{t\geq 0}$ on the domain\footnote{
The method can be adapted to other commonly seen domains without much
difficulty. We choose this particular domain since most asset prices are
positive.} ${\Bbb R}_{+}$. \ We focus on representing the transition
probability density function of $X_{t}$ in the one-dimensional case, $
f(y\mid x):=P(X_{t+h }=y\mid X_{t}=x)$. \

Our new method starts with a representation of the transition density using
Dirac's Delta function stated in the following Lemma.

\begin{lemma}
We have the following representation of the
transition density function:
\begin{align}
f(y\mid x)&={\Bbb E}_x[\delta (X_{t+h }-y)],  \label{step1}
\end{align}
where $\delta (\cdot )$ is the Dirac Delta function.
\end{lemma}
\pf From the sifting property of the Dirac Delta
function, we have the following calculation:
\[
{\Bbb E}_{x}[\delta (X_{t+h }-y)]=\int_{{\Bbb R}_{+}}\delta (u-y)f(u\mid
x)du=f(y\mid x).
\] This completes the proof.
\eproof

The above representation \eqref{step1} has also been utilized in the
literature. \ For example, \cite{Yang2019} used it to develop an
approximate parametric transition density expansion by approximating the
Dirac Delta function through a sequence of Gaussian functions. \ In contrast, the second step of our method involves using concrete and exact
representations of the Dirac Delta function from complete orthonomal basis
of Hilbert spaces.  To fix ideas, let $\left\{ g_{k}\left( y\right)
\right\} _{k=0}^{\infty }$ be a complete orthonomal basis, then the Dirac
Delta function can be represented by
\begin{equation}
\delta (x-a)=\sum_{k=0}^{\infty }g_{k}(x)g_{k}\left( a\right) .
\label{Delta_Generic}
\end{equation}

Note that the above identity holds in a \textquotedblleft distribution"
sense in the notation of generalized functions, and that the above representation is
essentially equivalent to the completeness of the basis. Thus the above
identity is also called the \textquotedblleft completeness identity".
What is more, there are many choices of the basis for various situations.
For more details, we refer the reader to an excellent expository article \citep{Li2013}. For example, we have the following series representations\footnote{The following formulas appear respectively as  formula (1.17.22), (1.17.24) and (1.17.23) on page 38 of the NIST handbook of Mathematical Functions, a definite reference on special functions. See \url{https://dlmf.nist.gov}} of
the Dirac Delta function:
\begin{align}
\delta (x-a)& =\sum\limits_{k=0}^{\infty }\left( k+\frac{1}{2}\right)
P_{k}(x)P_{k}(a),\   \nonumber \\
\delta (x-a)& =\frac{e^{-(x^{2}+a^{2})/2}}{\sqrt{\pi }}\sum\limits_{k=0}^{%
\infty }\frac{1}{2^{k}k!}H_{k}(x)H_{k}(a),\   \nonumber \\
\delta (x-a)& =e^{-(x+a)/2}\sum\limits_{k=0}^{\infty }L_{k}(x)L_{k}(a),\nonumber
\end{align}%
and each of the above respectively corresponds to the choices of $%
g_{k}(x)=\left( k+\frac{1}{2}\right) ^{1/2}P_{k}(x)$, $g_{k}(x)=\frac{1}{\pi
^{1/4}2^{k/2}\sqrt{k!}}e^{-x^{2}/2}H_{k}(x)$, and $g_{k}(x)=e^{-x/2}L_{k}(x)$%
. Here $P_{k}(x)$, $H_{k}(x)$ and $L_{k}(x)$ are respectively the Legendre,
Hermite and Laguerre polynomials, all of which are representative orthogonal polynomials. These series representations also appear
in  Theorem 2 on page 71 of \cite{Lebedev1965}. A proof based on asymptotic
behaviors of the corresponding special functions can be found, for example,
in \citep{Li2013}.

Our method is valid for the above three representations as well as others.
To keep the discussion general, in the following we use the more generic
representation \eqref{Delta_Generic}.  Combining \eqref{step1} and
\eqref{Delta_Generic}, we obtain the following representation of the transition
density function:
\begin{align}\label{step2}
f(y\mid x)& ={\Bbb E}_{x}[\delta (X_{t+h }-y)] ={\Bbb E}_{x}\left[
\sum_{k=0}^{\infty }g_{k}(X_{t+h })g_{k}\left( y\right) \right]
=\sum_{k=0}^{\infty }g_{k}\left( y\right) {\Bbb E}_{x}[g_{k}(X_{t+h })].
\end{align}

In the mult-dimensional setup, consider a multi-dimensional Markov process $%
{\bf X}_{t}:=(X_{t}^{(1)},X_{t}^{(2)},\cdots ,X_{t}^{(n)})$. \ We denote its
its transition density by:
\[
f({\bf y}\mid {\bf x}):=P\left( {\bf X}_{t+h }=%
{\bf y}\mid {\bf X}_{t}={\bf x}\right) ,
\]%
where ${\bf y}:=(y_{1},y_{2},\cdots ,y_{n})$ and ${\bf x}:=(x_{1},x_{2},%
\cdots ,x_{n})$. \ It should be noted that we allow for the components of
the multi-dimensional stochastic process ${\bf X}_{t}$ to have non-zero
correlations. We achieve the extension of our method by extending each of
the key representations in the one-dimensional case. The following Lemma
extends the equation \eqref{step1} in Lemma 1 to the multi-dimensional case.

\begin{lemma}
We have the following representation of the
d-dimensional transition density function:
\begin{equation}
f({\bf y}\mid {\bf x})={\Bbb E}_{{\bf x}}\left[ \prod\limits_{i=1}^{d}\delta
(X_{t+h }^{(i)}-y_{i})\right] ,  \label{stepm1}
\end{equation}%
where $\delta (\cdot )$ is the Dirac Delta function.
\end{lemma}
\pf
 From the sifting property of the Dirac Delta
function, we have the following calculations:
\[
{\Bbb E}_{{\bf x}}\left[ \prod\limits_{i=1}^{d}\delta (X_{t+h
}^{(i)}-y_{i})\right] =\int_{{\Bbb R}^{n}}\prod\limits_{i=1}^{d}\delta (u_{i}-y_{i})\cdot f({\bf u}\mid {\bf x}
)d{\bf u}= f({\bf y}\mid {\bf x}),
\]
where in the second equality is obtained from
the sifting property of the Dirac Delta function. This completes the proof.
\eproof

\begin{remark}
Note that in \eqref {stepm1}, we just need the one-dimensional Dirac Delta
function. The intuitive reason can be seen from the above proof: as we
gradually integrate out the dummy variables one by one while freezing the
other dummy variables. This turns out to be convenient for us to provide an
explicit representation of the multi-variate transition density function of
the multi-dimensional Markov process.
\end{remark}

Now we move on to extend the representation \eqref {step2} to the multi-dimensional case.
Substituting the representation (\ref{Delta_Generic}) for each Dirac Delta
function into equation \eqref {stepm1}, we obtain the following explicit
representation of the transition density function of the multi-dimensional
Markov process ${\bf X}$:
\begin{align}
f({\bf y}\mid {\bf x})& ={\Bbb E}_{{\bf x}}\left[ \prod\limits_{i=1}^{d}
\delta (X_{t+h }^{(i)}-y_{i})\right] ={\Bbb E}_{{\bf x}}\left[
\prod\limits_{i=1}^{d}\left( \sum\limits_{k_{i}=0}^{\infty
}g_{k_{i}}^{(i)}(X_{t+h }^{(i)})g_{k_{i}}^{(i)}(y_{i})\right) \right]
\nonumber \\
& =\sum\limits_{k_{1}, \ldots, k_{n}=0}^{\infty }\left( \prod\limits_{i=1}^{d}g_{k_{i}}^{(i)}(y_{i})\right) {\Bbb E}_{{\bf x}
}\left[ \prod\limits_{i=1}^{d}g_{k_{i}}^{(i)}(X_{t+h }^{(i)})\right].
\label{key-multi}
\end{align}

\subsection{Delta family algorithm for the stochastic control problems}\label{sec:multi-D}

Consider a general 1-D stochastic control problem with its value function characterized by
\begin{align}
V(t,x)&=\sup\limits_{\pi\in\mathcal{A}} {\Bbb E}[U(X_T)\mid X_t=x],\notag
\end{align}
where $U$ represents the utility function that is continuous on $[0,+\infty$) and satisfies $U(x)=-\infty$ for $x<0$, and $U(x)\leq{C}(1+x^{\varrho})$ for some constants $C>0$ and $0<\varrho<1$.
The process $X_t$ depends on the unknown control $\pi$.
From Bellman's dynamical programming principle, we have the following characterization of the value function:
\begin{align}
V(t,x)&=\sup\limits_{\pi\in\mathcal{A}} {\Bbb E}[V(t+h, X_{t+h})\mid X_t=x].\label{dpp}
\end{align}

Based on \eqref{dpp}, the traditional approach proceeds as follows:
\begin{enumerate}
\item Assume $V(t,x)\in C^{1,2}$, and then apply the Ito's lemma to rewrite the right hand side of \eqref{dpp} into a differential form.
\item Then take the limit $h\rightarrow 0$ to characterize the local behavior of the value function. This is achieved through obtaining the HJB PDE, which is a highly nonlinear PDE and challenging to solve.
\end{enumerate}

We shall take an alternative approach, and the idea is to rewrite the conditional expectation in \eqref{dpp} using the conditional transition density function, and then use our previous representation in \eqref{step2} to represent the density function. Note that the unknown control is only associated with the conditional density function.
We have
\begin{align}
{\Bbb E}[V(t+h, X_{t+h})\mid X_t=x]&=\int_{\mathbb R_+} V(t+h, y) f(y\mid x) dy\notag\\
&=\sum_{k=0}^{\infty} \left(\int_{\mathbb R_+} V(t+h, y)g_{k}\left( y\right) dy\right) {\Bbb E}_{x}[g_{k}(X_{t+h })]\notag\\
&=\sum_{k=0}^{\infty} \left(\int_{\mathbb R_+} V(t+h, y)g_{k}\left( y\right) dy\right)  \left(\int_t^{t+h}  {\Bbb E}_{x}[{\mathcal L}^{\pi} g_{k}(X_{u})] du+g_{k}(x)\right),\label{new-rep}
\end{align}
where ${\mathcal L}^{\pi}$ is the infinitesimal generator of the Markov process $X$, and note that it contains the unknown control.
Plugging \eqref{new-rep} into \eqref{dpp}, we have
\begin{align}\label{dpp2}
&V(t,x)=\sup\limits_{\pi\in\mathcal{A}} {\Bbb E}[V(t+h, X_{t+h})\mid X_t=x]\notag\\
&=\sup\limits_{\pi\in\mathcal{A}}  \sum_{k=0}^{\infty} \left(\int_{\mathbb R_+} V(t+h, y)g_{k}\left( y\right) dy\right)  \left(\int_t^{t+h}  {\Bbb E}_{x}[{\mathcal L}^{\pi} g_{k}(X_{u})] du+g_{k}(x)\right).
\end{align}
Assume that $h\rightarrow 0$ is very small, from the mean value theorem, we have the following approximation:  $\pi_u\rightarrow \pi_t=\pi$, which is an unknown constant, and $X_u\rightarrow X_t=x$. To summarize
\begin{align}
\int_t^{t+h}  {\Bbb E}_{x}[{\mathcal L}^{\pi} g_{k}(X_{u})] du&\approx h \cdot{\mathcal L}^{\pi} g_k (x). \label{approx1}
\end{align}
Combining \eqref{dpp2} and \eqref{approx1}, we have
\begin{align}\label{dpp3}
&V(t,x)\approx \sup\limits_{\pi\in\mathcal{A}}  \sum_{k=0}^{\infty} \left(\int_{\mathbb R_+} V(t+h, y)g_{k}\left( y\right) dy\right) \left(h \cdot {\mathcal L}^{\pi} g_k (x)+g_{k}(x)\right).
\end{align}
The terminal condition is given by
\begin{equation}\label{terminal}
V(T,x)=U(x),\notag
\end{equation}
where $U(\cdot)$ is the known utility function.

Let $\pi^{*}$ be the solution of
\begin{equation}\label{1st-order-condition-1D}
 \sum_{k=0}^{\infty} \left(\int_{\mathbb R_+} V(t+h, y)g_{k}\left( y\right) dy\right) \frac{\partial {\mathcal L}^{\pi} g_k (x)}{\partial \pi} =0.
\end{equation}
Then inserting $\pi=\pi^*$ into \eqref{dpp3} gives that
\begin{align}\label{dpp3-HJB}
&V(t,x)\approx  \sum_{k=0}^{\infty} \left(\int_{\mathbb R_+} V(t+h, y)g_{k}\left( y\right) dy\right) \left(h \cdot {\mathcal L}^{\pi^*} g_k (x)+g_{k}(x)\right).
\end{align}
Taking inner product to \eqref{dpp3-HJB} with $g_{j}(x),\; j=0,1,\ldots$ and denoting
\[
V_{j}(t):=\int_{\mathbb R_+}V(t,x)g_{j}(x)dx,\quad j=0,1,\ldots,
\]
we obtain that
\begin{align}\label{dpp5-HJB}
V_{j}(t)
&\approx V_{j}(t+h)+h\sum_{k=0}^{\infty} V_{k}(t+h) \int_{\mathbb R_+}g_{j}(x) {\mathcal L}^{\pi^{*}} g_k (x)dx,\quad j=0,1,\ldots,
\end{align}
with $\pi^*$ being the solution of \eqref{1st-order-condition-1D} which can be rewritten as
\begin{align}\label{1st-order-condition-1D-new-form}
 \sum_{k=0}^{\infty} V_{k}(t+h) \frac{\partial {\mathcal L}^{\pi} g_k (x)}{\partial \pi} =0.
\end{align}
Using the orthogonality of the basis functions $g_{j}(x)$, we can obtain that
\begin{equation}\label{representation-value-function-1D}
 V(t,x)= \sum_{j=0}^{\infty}V_{j}(t)g_{j}(x).
\end{equation}
Truncating the sums in \eqref{dpp5-HJB}, \eqref{1st-order-condition-1D-new-form} and \eqref{representation-value-function-1D} by a finite term $M$ leads to the following time-stepping algorithm.

\begin{algorithm}[H]
\caption{(Delta family algorithm for 1-D stochastic control problems)}\label{alg:1-D}
\begin{algorithmic}[1]
\STATE  We shall divide the interval $[0,T]$ into $N$ equal sub-intervals at a time step size of $h:=T/N$. Denote $t_{n}:=nh$ for $n=0,1,\ldots, N$ and $V^{n}_{j}\approx V_{j}(t_{n})$ for $j=0,1,\ldots, M$.
\STATE  The algorithm is given by the following recursion for $n=0,1,\ldots,N-1$,
\begin{align}\label{dpp5-algorithm-1D}
V^{n}_{j}
&= V^{n+1}_{j}+h\sum_{k=0}^{M} V^{n+1}_{k}  \int_{\mathbb R_+} g_{j}(x){\mathcal L}^{\pi^{*}_{n+1}} g_k (x)dx,\quad j=0,1,\ldots, M,
\end{align}
with $\pi^{*}_{n+1}$ being the solution of
\begin{align}\label{1st-order-condition-1D-new-form-discrete}
 \sum_{k=0}^{\infty} V_{k}^{n+1} \frac{\partial {\mathcal L}^{\pi} g_k (x)}{\partial \pi} =0,\notag
\end{align}
and terminal condition
\begin{equation}\label{terminal-condition-1D}
V^{N}_{j}=\int_{\mathbb R_+}U(x)g_{j}(x)dx,\quad j=0,1,\ldots, M.
\end{equation}
The integrals in \eqref{dpp5-algorithm-1D} and \eqref{terminal-condition-1D} can be evaluated by quadrature rules.
\STATE Finally from \eqref{representation-value-function-1D}, we obtain the approximation of the value functions
\begin{equation}\label{approx-value-function-1D}
 V(t_{n},x)\approx \sum_{j=0}^{\infty}V^{n}_{j}g_{j}(x),\quad n=0,1,\ldots,N,\notag
\end{equation}
and the optimal strategies
\begin{equation}\label{strategy-approx-1D}
\pi^{*}(t_{n},x)\approx \pi^{*}_{n}(x),\quad n=0,1,\ldots,N.\notag
\end{equation}
\end{algorithmic}
\end{algorithm}

\newpage

For the high-dimensional Markov process, consider the following generic stochastic control problem:
Consider a general stochastic control problem with its value function characterized by
\begin{align}
V(t, {\bf x})&=\sup\limits_{{\pi}\in\mathcal{A}} {\Bbb E}[U({\bf X}_T)\mid {\bf X_t=x}].\notag
\end{align}
From Bellman's dynamical programming principle, we have the following characterization of the value function:
\begin{align}
V(t,{\bf x})&=\sup\limits_{{\pi}\in\mathcal{A}} {\Bbb E}[V(t+h, {\bf X}_{t+h})\mid {\bf X_t=x}].\label{dpp-n}
\end{align}
Recall that intuitively the dynamical programming principle is a backward valuation in the time space. Note that the time space is always one dimensional, although the state space of the underlying Markov process can be of high dimensions. We shall utilize the density representation in \eqref{key-multi} to rewrite the conditional expectations on the right hand side of \eqref{dpp-n} as follows.
\begin{align}
&{\Bbb E}[V(t+h, {\bf X}_{t+h})\mid {\bf X_t=x}]=\int_{\mathbb R^n_+} V(t+h, {\bf y}) f({\bf y}\mid {\bf x}) d{\bf y}\notag\\
&\quad =\sum\limits_{k_{1}, \ldots, k_{n}=0}^{\infty }\left( \int_{\mathbb R^n_+} V(t+h, {\bf y}) \prod\limits_{i=1}^{d}g_{k_{i}}^{(i)}(y_{i})d{\bf y}\right) {\Bbb E}_{{\bf x}
}\left[ \prod\limits_{i=1}^{d}g_{k_{i}}^{(i)}(X_{t+h }^{(i)})\right]\notag\\
&\quad =\sum\limits_{k_{1}, \ldots, k_{n}=0}^{\infty }\left( \int_{\mathbb R^n_+} V(t+h, {\bf y}) \prod\limits_{i=1}^{d}g_{k_{i}}^{(i)}(y_{i})d{\bf y}\right) \left( \prod\limits_{i=1}^{d}g_{k_{i}}^{i}(x^{i}) + \int_t^{t+h} {\Bbb E}_{{\bf x}
}\left[ {\mathcal L}^{\pi}\left(\prod\limits_{i=1}^{d}g_{k_{i}}^{(i)}(X_{u}^{(i)})\right)\right]du \right),\notag
\end{align}
where ${\mathcal L}^{\pi}$ is the infinitesimal generator of the multi-dimensional Markov process ${\bf X}$.
Note that we can still use the one-dimensional mean value theorem in the time space, and arrive at the following first order approximation:
\begin{align}
\int_t^{t+h} {\Bbb E}_{{\bf x}}\left[ {\mathcal L}^{\pi}\left(\prod\limits_{i=1}^{d}g_{k_{i}}^{(i)}(X_{u}^{(i)})\right)\right]du&\approx h\cdot {\mathcal L}^{\pi}\left(\prod\limits_{i=1}^{d}g_{k_{i}}^{(i)}(x_{i})\right).\notag
\end{align}
Then we finally reduce the problem into solving the  problem below, similar as the one-dimensional case,
\begin{align}
V(t, {\bf x})&=\sup\limits_{{\pi}\in\mathcal{A}} \sum\limits_{k_{1}, \ldots, k_{n}=0}^{\infty }\left( \int_{\mathbb R^n_+} V(t+h, {\bf y}) \prod\limits_{i=1}^{d}g_{k_{i}}^{(i)}(y_{i})d{\bf y}\right) \int_t^{t+h} {\Bbb E}_{{\bf x}
}\left[ {\mathcal L}^{\pi}\left(\prod\limits_{i=1}^{d}g_{k_{i}}^{(i)}(X_{u}^{(i)})\right)\right]du\notag\\
&\approx \sup\limits_{{\pi}\in\mathcal{A}} \sum\limits_{k_{1}, \ldots, k_{n}=0}^{\infty }\left( \int_{\mathbb R^n_+} V(t+h, {\bf y}) \prod\limits_{i=1}^{d}g_{k_{i}}^{(i)}(y_{i})d{\bf y}\right) \left(h\cdot {\mathcal L}^{\pi}\left(\prod\limits_{i=1}^{d}g_{k_{i}}^{(i)}(x_i)\right) + \prod\limits_{i=1}^{d}g_{k_{i}}^{(i)}(x_{i}) \right).\label{multi-app}
\end{align}
Let $\pi^{*}$ be the solution of
\begin{equation}\label{1st-order-condition-multi-D}
 \sum_{k_1,\ldots,k_n=0}^{\infty} \left(\int_{\mathbb R^n_+}  V(t+h, {\bf y}) \prod\limits_{i=1}^{d}g_{k_{i}}^{(i)}(y_{i})d{\bf y}\right) \frac{\partial}{\partial \pi} {\mathcal L}^{\pi}\left(\prod\limits_{i=1}^{d}g_{k_{i}}^{(i)}(x_i)\right) =0.
\end{equation}
Then inserting $\pi=\pi^*$ into \eqref{multi-app} gives that
\begin{align}\label{multi-app-HJB}
V(t, {\bf x})
&\approx  \sum\limits_{k_{1}, \ldots, k_{n}=0}^{\infty }\left( \int_{\mathbb R^n_+} V(t+h, {\bf y}) \prod\limits_{i=1}^{d}g_{k_{i}}^{(i)}(y_{i})d{\bf y}\right) \left( h\cdot {\mathcal L}^{\pi^{*}}\left(\prod\limits_{i=1}^{d}g_{k_{i}}^{(i)}(x_i)\right) + \prod\limits_{i=1}^{d}g_{k_{i}}^{(i)}(x_{i}) \right).\notag
\end{align}
Taking inner product to \eqref{dpp3-HJB} with $\prod\limits_{j=1}^{d}g_{m_{j}}^{(j)}(x_j),\; m_j=0,1,\ldots;\, j=1,\ldots,n$ and denoting
\[
V_{\bf m}(t):=\int_{{\mathbb R}^{n}_+}V(t,{\bf x}) \prod\limits_{j=1}^{d}g_{m_{j}}^{(j)}(x_j)d{\bf x},\quad m_j=0,1,\ldots;\, j=1,\ldots,n,
\]
we obtain that for $m_j=0,1,\ldots;\, j=1,\ldots,n$,
\begin{align}\label{dpp5-HJB-multi-D}
V_{{\bf m}}(t)
&\approx V_{{\bf m}}(t+h) + h\sum\limits_{k_{1}, \ldots, k_{n}=0}^{\infty } V_{{\bf k}}(t+h) \int_{{\mathbb R}^{n}_+}\prod\limits_{j=1}^{d}g_{m_{j}}^{(j)}(x_j) {\mathcal L}^{\pi^{*}}\left(\prod\limits_{i=1}^{d}g_{k_{i}}^{(i)}(x_i)\right)d{\bf x},
\end{align}
with $\pi^*$ being the solution of \eqref{1st-order-condition-multi-D} which can be rewritten as
\begin{align}\label{1st-order-condition-multi-D-new-form}
 &\sum_{k_1,\ldots,k_n=0}^{\infty} V_{{\bf k}}(t+h)\frac{\partial}{\partial \pi} {\mathcal L}^{\pi}\left(\prod\limits_{i=1}^{d}g_{k_{i}}^{(i)}(x_i)\right) =0.
\end{align}
Using the orthogonality of the basis functions $g_{j}(x)$, we can obtain that
\begin{equation}\label{representation-value-function-multi-D}
 V(t,{\bf x})= \sum_{k_1,\ldots,k_n=0}^{\infty} V_{{\bf k}}(t)\prod\limits_{i=1}^{d}g_{k_{i}}^{(i)}(x_i).
\end{equation}
Truncating the sums in \eqref{dpp5-HJB-multi-D}, \eqref{1st-order-condition-multi-D-new-form} and \eqref{representation-value-function-multi-D} by finite number leads to the following time-stepping algorithm.

\begin{algorithm}[H]
\caption{(Delta family algorithm for multi-dimensional stochastic control problems)}\label{alg:multi-D}
\begin{algorithmic}[1]
\STATE  Denote $V^{n}_{{\bf m}}\approx V_{{\bf m}}(t_{n})$ for ${\bf m}=(m_1,\ldots,m_n)$;
$m_j=0,1,\ldots, M_j$; $j=1,\ldots, n$; $n=0,1,\ldots, N$.
\STATE  The algorithm is given by the following recursion for $n=0,1,\ldots,N-1$,
\begin{align}\label{value-multi-D-algorithm}
V_{{\bf m}}^{n}&\approx V_{{\bf m}}^{n+1} + h\sum\limits_{k_{1}=0}^{M_1 }\cdots\sum\limits_{k_{n}=0}^{M_n}
 V_{{\bf k}}^{n+1} \int_{{\mathbb R}^{n}_+}\prod\limits_{j=1}^{d}g_{m_{j}}^{(j)}(x_j) {\mathcal L}^{\pi^{*}_{n+1}}\left(\prod\limits_{i=1}^{d}g_{k_{i}}^{(i)}(x_i)\right)d{\bf x},
\end{align}
with $\pi^*_{n+1}$ being the solution of \eqref{1st-order-condition-multi-D} which can be rewritten as
\begin{align}
 &\sum\limits_{k_{1}=0}^{M_1 }\cdots\sum\limits_{k_{n}=0}^{M_n} V_{{\bf k}}^{n+1}\frac{\partial}{\partial \pi} {\mathcal L}^{\pi}\left(\prod\limits_{i=1}^{d}g_{k_{i}}^{(i)}(x_i)\right) =0.\notag
\end{align}
The terminal condition
\begin{equation}\label{terminal-condition-multi-D}
V^{N}_{{\bf m}}=\int_{\mathbb R_+^n}U({\bf x})\prod\limits_{j=1}^{d}g_{m_{j}}^{(j)}(x_j)d {\bf x}.
\end{equation}
The integrals in \eqref{value-multi-D-algorithm} and \eqref{terminal-condition-multi-D} can be evaluated by quadrature rules.
\STATE Finally from \eqref{representation-value-function-multi-D}, we obtain the approximation of the value functions
\begin{equation}\label{approx-value-function-multi-D}
 V(t_{n},{\bf x})\approx \sum\limits_{m_{1}=0}^{M_1 }\cdots\sum\limits_{m_{n}=0}^{M_n} V_{{\bf m}}^{n}\prod\limits_{j=1}^{d}g_{m_{j}}^{(j)}(x_j),\notag
\end{equation}
and the optimal strategies
\begin{align}
\pi^{*}(t_{n},x)&\approx \pi^{*}_{n}(x),\quad n=0,1,\ldots,N.\notag
\end{align}
\end{algorithmic}
\end{algorithm}

\section{Applications and examples}\label{sec3}
In this section, we apply the delta family algorithms to solve the optimal investment problems under the Heston model, the optimal reinsurance and investment problems under respectively the Heston model and the rough Heston model, and optimal stopping investment problems under general stochastic local volatility (SLV) models

\subsection{Optimal investment under the Heston model}
Consider the Heston model:
\begin{align}\label{heston-S}
\frac{dS_t}{S_t}&=(r+\lambda \mathcal{V}_t) dt +\sqrt{\mathcal{V}_t} dW_t^{(1)},\\
\label{heston-V}
d\mathcal{V}_t&=\kappa(\theta-\mathcal{V}_t) dt +\sigma \sqrt{\mathcal{V}_t}dW_t^{(2)},
\end{align}
where $E[dW_t^{(1)} dW_t^{(2)}]=\rho dt$.
By definition, the wealth process is given by
\begin{align}\label{heston-wealth}
\frac{dX_t}{X_t}&=\pi_t \frac{dS_t}{S_t} +(1-\pi_t) rdt=\lambda\pi_t\mathcal{V}_t dt +\pi_t \sqrt{\mathcal{V}_t} dW_t^{(1)},
\end{align}
where $\mathcal{V}$ is given by \eqref{heston-V}.
The infinitesimal generator of \eqref{heston-wealth} is given by
\begin{align}\label{IG-hes}
{\mathcal L}^{\pi} f(x,v) = \lambda\pi vx\frac{\partial f}{\partial x}+\kappa(\theta-v) \frac{\partial f}{\partial v}+\frac{\pi^2 vx^2}{2}  \frac{\partial^2 f}{\partial x^2} +\frac{\sigma^2 v}{2} \frac{\partial^2 f}{\partial v^2}+ \rho \pi \sigma x v  \frac{\partial^2 f}{\partial x \partial v}.
\end{align}
A key observation is that the above infinitesimal generator is still a quadratic function in $\pi$. This is very convenient in working out the explicit optimizers, as seen in the one-dimensional case. To use the Legendre functions and the Gauss-Legendre quadrature rules, we set $(x,v)\in[x_{\min},x_{\max}]\times [v_{\min},v_{\max}]$ and transform it into $(y_1, y_2)\in\Omega^{2}:=[-1,1]^{2}$ using the following transformation
\begin{align}
\hat{x}(y_1) &= (x_{\max}-x_{\min})(y_1+1)/2+x_{\min},\label{hat_x}\\
\hat{v}(y_2) &= (v_{\max}-v_{\min})(y_2+1)/2+v_{\min},\\
\tilde{x}(x) &= \frac{2(x-x_{\min})}{x_{\max}-x_{\min}}-1,\\
\tilde{v}(v) &= \frac{2(v-v_{\min})}{v_{\max}-v_{\min}}-1.\label{tilde_v}
\end{align}
By combining \eqref{multi-app} and \eqref{IG-hes}, we obtain that
\begin{align}\label{value-Heston-1}
V(t,x,v)\approx\sup\limits_{{\pi}\in\mathcal{A}} \sum\limits_{k_{1}, k_2=0}^{\infty }
&\left( \int_{\Omega^{2}} V(t+h,\hat{x}(y_1),\hat{v}(y_2)) g_{k_{1}}^{(1)}(y_1) g_{k_{2}}^{(2)}(y_2) d{\bf y}\right)  \notag\\ &\cdot\left(g_{k_{1}}^{(1)}(\tilde{x}(x)) g_{k_{2}}^{(2)}(\tilde{v}(v)) + h\cdot {\mathcal L}^{\pi}\left(g_{k_{1}}^{(1)}(\tilde{x}(x)) g_{k_{2}}^{(2)}(\tilde{v}(v))\right) \right),
\end{align}
where $\mathbf{y} = (y_1,y_2) $, and
\begin{align}
&{\mathcal L}^{\pi}\left(g_{k_{1}}^{(1)}(\tilde{x}(x)) g_{k_{2}}^{(2)}(\tilde{v}(v))\right)\notag\\
&= \lambda\pi vx\frac{d g_{k_{1}}^{(1)}(\tilde{x}(x)) }{d x} g_{k_{2}}^{(2)}(\tilde{v}(v))+\kappa(\theta-v) g_{k_{1}}^{(1)}(\tilde{x}(x))\frac{d g_{k_{2}}^{(2)}(\tilde{v}(v))}{d v}+\frac{\pi^2 vx^2}{2}  \frac{d^2 g_{k_{1}}^{(1)}(\tilde{x}(x))}{d x^2}g_{k_{2}}^{(2)}(\tilde{v}(v))\notag\\
& \quad +\frac{\sigma^2 v}{2}g_{k_{1}}^{(1)}(\tilde{x}(x)) \frac{d^2 g_{k_{2}}^{(2)}(\tilde{v}(v))}{d v^2}+ \rho \pi \sigma x v  \frac{d g_{k_{1}}^{(1)}(\tilde{x}(x)) }{d x}\frac{d g_{k_{2}}^{(2)}(\tilde{v}(v))}{d v}.\notag
\end{align}
The first order condition gives that
\begin{align}\label{strategy-Heston-1}
&\pi^{*}=\\
&\frac{-\sum\limits_{k_{1}, k_2=0}^{\infty }
\left( \int_{\Omega^{2}} V(t+h, \hat{x}(y_1),\hat{v}(y_2)) g_{k_{1}}^{(1)}(y_{1}) g_{k_{2}}^{(2)}(y_{2}) d{\bf y}\right)
\left(\lambda g_{k_{2}}^{(2)}(\tilde{v}(v))+\rho \sigma \frac{d g_{k_{2}}^{(2)}(\tilde{v}(v))}{d v}\right)\frac{d g_{k_{1}}^{(1)}(\tilde{x}(x))}{d x}}{x\sum\limits_{k_{1}, k_2=0}^{\infty }
\left( \int_{\Omega^{2}} V(t+h, \hat{x}(y_1),\hat{v}(y_2)) g_{k_{1}}^{(1)}(y_{1}) g_{k_{2}}^{(2)}(y_{2}) d{\bf y}\right)\frac{d^{2}g_{k_{1}}^{(1)}(\tilde{x}(x))}{d x^{2}} g_{k_{2}}^{(2)}(\tilde{v}(v))}.\notag
\end{align}
Inserting \eqref{strategy-Heston-1} into \eqref{value-Heston-1} gives that
\begin{align}\label{value-Heston-2}
V(t,x,v)\approx \sum\limits_{k_{1}, k_2=0}^{\infty }
&\left( \int_{\Omega^{2}} V(t+h,\hat{x}(y_1),\hat{v}(y_2)) g_{k_{1}}^{(1)}(y_1) g_{k_{2}}^{(2)}(y_2) d{\bf y}\right)  \notag\\ &\cdot\left(g_{k_{1}}^{(1)}(\tilde{x}(x)) g_{k_{2}}^{(2)}(\tilde{v}(v)) + h\cdot {\mathcal L}^{\pi^{\ast}}\left(g_{k_{1}}^{(1)}(\tilde{x}(x)) g_{k_{2}}^{(2)}(\tilde{v}(v))\right) \right).\notag
\end{align}
Denote
\[
V_{m_1 m_2}(t):= \int_{\Omega^{2}} V(t,\hat{x}(y_1),\hat{x}(y_2)) g_{m_{1}}^{(1)}(y_1) g_{m_{2}}^{(2)}(y_2) d\bf{y}.
\]
Then the time-stepping algorithm as the corollary of Algorithm~\ref{alg:2-D} is given as follows.
\begin{algorithm}[H]
\caption{(Optimal investment under Heston models)}\label{alg:2-D}
\begin{algorithmic}[1]
\STATE  Denote $V^{n}_{m_1 m_2}\approx V_{m_1 m_2}(t_{n})$ for $m_1=0,1,\ldots, M_1$; $m_2=0,1,\ldots, M_2$; $n=0,1,\ldots, N$.
\STATE  The algorithm is given by the following recursion for $n=0,1,\ldots,N-1$,
\begin{align}\label{value-Heston-2-algorithm}
V_{m_1m_2}^{n}&\approx V_{m_1m_2}^{n+1} + h\sum\limits_{k_{1}=0}^{M_1 }\sum\limits_{k_{2}=0}^{M_2}
 V_{k_1k_2}^{n+1} \int_{\Omega^{2}} g_{m_{1}}^{(1)}(y_1) g_{m_{2}}^{(2)}(y_2){\mathcal L}^{\bf \pi^{*}_{n+1}}\left(g_{k_{1}}^{(1)}(y_1) g_{k_{2}}^{(2)}(y_2)\right)d\bf{y},
\end{align}
where
\begin{align}\label{strategy-Heston-2-algorithm}
\pi^{*}_{n+1}&=\frac{-\sum\limits_{k_{1}=0}^{M_1}\sum\limits_{k_{2}=0}^{M_2}
V_{k_1k_2}^{n+1} \left(\lambda g_{k_{2}}^{(2)}(\hat{v}(v))+\rho \sigma \frac{d g_{k_{2}}^{(2)}(\hat{v}(v))}{d v}\right)\frac{d g_{k_{1}}^{(1)}(\hat{x}(x))}{d x}}{x\sum\limits_{k_{1}=0}^{M_1 }\sum\limits_{k_{2}=0}^{M_2}
V_{k_1k_2}^{n+1}\frac{d^{2}g_{k_{1}}^{(1)}(\hat{x}(x))}{d x^{2}} g_{k_{2}}^{(2)}(\hat{v}(v))},\notag
\end{align}
and terminal condition
\begin{equation}\label{terminal-condition-Heston}
V^{N}_{m_1m_2}=\int_{\Omega^2}U(\hat{x}(y_1))g_{m_1}^{(1)}(\hat{x}(y_1))g_{m_2}^{(2)}(\hat{v}(y_2))d\mathbf{y},\quad m_1=0,1,\ldots, M_1;\, m_2=0,1,\ldots,M_2.
\end{equation}
The integrals in \eqref{value-Heston-2-algorithm} and \eqref{terminal-condition-Heston} are evaluated by Gauss-Legendre quadrature rules.
\STATE Finally  we obtain the approximation of the value functions
\begin{equation}\label{approx-value-function-Heston}
 V(t_{n},x,v)\approx \sum\limits_{m_{1}=0}^{M_1 }\sum\limits_{m_{2}=0}^{M_2} V_{m_1m_2}^{n}g_{m_1}^{(1)}(\hat{x}(x))g_{m_2}^{(2)}(\hat{v}(v)),\notag
\end{equation}
for $m_1=0,1,\ldots, M_1;\, m_2=0,1,\ldots, M_2;\, n=0,1,\ldots, N$
and the optimal strategies
\begin{align}\label{strategy-approx-Heston}
\pi^{*}(t_{n},x,v)&\approx \pi^{*}_{n}(x,v).\notag
\end{align}
\end{algorithmic}
\end{algorithm}

\begin{exam}\label{exam:HestonPower}
In this example, we solve  the utility maximization problem under the Heston model with power utility $U(x)=\frac{x^{1/2}}{1/2}$ and
with parameters taken from \cite{Kraft2005}
\begin{equation}
r=0.05,\, \rho=-0.5,\, \kappa=10,\, \theta=0.05,\, \sigma=0.5,\, \lambda=0.5,\, \mathcal{V}_0=0.5,\, T=1.\notag
\end{equation}
We compute the errors of the prime value and strategy at time $0$ using the maximum norm on the
range of the initial wealth and volatility $x\in [a,b]=[1,2]$ and $v\in [c,d]= [0.3,0.6]$. The benchmark values are give by \cite{Ma2020}.
\end{exam}
For the computation of the delta family approach, we take
$$x_{\min}=(3a-b)/2,\, x_{\max}=(9b-7a)/2,\, v_{\min}=(3c-d)/2,\, v_{\max}=(9d-7c)/2,$$
 namely in this example,
 $$x_{\min}=0.5,\, x_{\max}=5.5,\, v_{\min}=0.15,\, v_{\max}=1.65.$$
From the numerics in Table~\ref{Table:Heston-err}, it is observed that the errors are decreasing with the number of basis functions $M$ increasing. In the Figure~\ref{Figure:Heston-logerr},
we compare computational efficiency between delta family approach and the dual control Monte-Carlo method in \cite{Ma2020}. It can be seen from the figure that the delta family method is much faster than the dual control Monte-Carlo method. In the implementation of Algorithm~\ref{alg:2-D},
the integrals in \eqref{value-Heston-2-algorithm} and \eqref{terminal-condition-Heston} can be evaluated by Gauss-Legendre quadrature rules with number of nodes $40$.

\begin{table}[!htbp]
\centering
\caption{Errors of delta family approach with number of time steps $N = 2000$ (for Example \ref{exam:HestonPower}).}
\begin{tabular}{c c c c c c c c}\\
\hline
 \multirow{2}{*}{$M$} & \multicolumn{2}{c}{Errors} &\multirow{2}{*}{CPU time (Seconds)}\\
\cline{2-3}
                 & Prime value    & Optimal strategy      \\
\cline{1-4}
6   &2.95e-4	&6.08e-2 	&2.729 \\
\cline{1-4}
8   &8.50e-5 	&9.46e-3 	&3.152 \\
\cline{1-4}
10  &2.97e-5 	&2.93e-3 	&3.956 \\
\cline{1-4}
12  &7.48e-6	&1.94e-3 	&7.356 \\
\cline{1-4}
14  &2.73e-6 	&1.12e-3 	&9.107 \\
\cline{1-4}
16  &1.31e-6 	&4.84e-4 	&14.294 \\
\hline
\end{tabular}
\label{Table:Heston-err}
\end{table}

\begin{figure}[!h]\label{Figure:Heston-logerr}
\centering
\includegraphics[height=6cm,width=8cm]{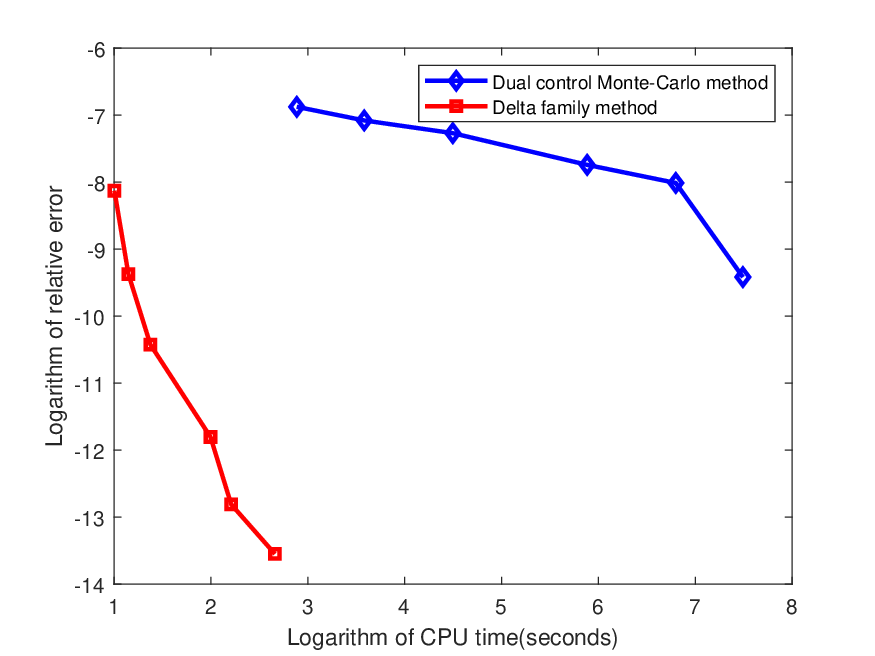}
\centering
\caption{Comparison between delta family approach and dual Monte-Carlo method (Example \ref{exam:HestonPower}).}
\label{Figure:Heston-logerr}
\end{figure}

\subsection{Optimal reinsurance-investment under (rough) Heston models}
Assume that the surplus process of the insurer satisfies (\cite{Bauerle-2005})
\begin{equation}
dR_t=c\left[\eta-\vartheta(1-\widehat{q}_t)\right]dt+b\widehat{q}_tdW_t,\notag
\end{equation}
where $c,\,b,\,\eta$ are positive constants representing the claim rate, volatility and safety loading of the insurer, respectively, and $W_t$ is the standard Brownian motion representing the uncertainty of the insurance market. The insurer chooses to participate in the reinsurance market to reduce the underlying risks or take an extra insurance business. $\widehat{q}_t$ is the proportion of reinsurance at time $t$, which means that the insurer undertakes $100\widehat{q}_t\%$ of the claims while the reinsurer undertakes the rest. Meanwhile, the insurer has to pay a premium at the rate of $c\vartheta(1-\widehat{q}_t)$ to the reinsurer due to the reinsurance business, where $\vartheta$ denotes the safety loading of reinsurer and $\vartheta >\eta$ is satisfied to exclude the arbitrage.
Further, the insurer has an option to invest in the continuous-time financial market that consists of one risky asset and one risk-free asset with interest $r>0$ to get more benefits. The risky asset price $S$ satisfies process \eqref{heston-S} and the variance $\mathcal{V}$ follows Heston model \eqref{heston-V} or rough Heston model
\begin{equation}\label{rough variance process}
\mathcal{V}_t=\mathcal{V}_0+\int^t_0K(t-s)\Big(\kappa(\theta-\mathcal{V}_s)ds+\sigma\sqrt{\mathcal{V}_s}dW_s^{2}\Big),\notag
\end{equation}
where $K(t)=\frac{t^{\alpha-1}}{\Gamma(\alpha)}$ and $0<\alpha<1$.
The standard Brownian motion $W_t^1$ for the risky asset price $S$ and $W_t^2$ for the process of variance $\mathcal{V}$ are correlated with correlation coefficient $-1\leq\rho\leq 1$ and they are both independent of $W_t$ for the surplus process.

We denote the wealth process of the insurer at time $t$ by $\widehat{X}_t$. Assume that the amount of wealth (investment strategy) invested in the risky asset $S$ and the risk-free asset $B$ is $\widehat{\pi}_t$ and $\widehat{X}_t-\widehat{\pi}_t$, respectively. Consequently, the dynamic equation of the wealth process $\widehat{X}_t$ evolves according to
\begin{eqnarray}\label{wealth process}
d\widehat{X}_t
=r\widehat{X}_tdt+\lambda \widehat{\pi}_t\mathcal{V}_tdt +\widehat{\pi}_t\sqrt{\mathcal{V}_t}dW_{t}^{1}+c(\eta-\vartheta)dt+c\vartheta\widehat{q}_tdt+b\widehat{q}_tdW_t,
\end{eqnarray}
with $\widehat{X}_t=\widehat{x}$. Here $(\widehat{q}_t,\widehat{\pi}_t)$ denotes admissible reinsurance-investment strategy. A reinsurance-investment strategy $\widehat{u}_t$, ${t\in[0, T]}$ is said to be admissible if it satisfies that $(\widehat{q}_t,\widehat{\pi}_t)$ is progressively measurable in regards to $\mathcal{F}_t$ which satisfies $E\Big[\int_{0}^{T}b^2\widehat{q}^2_tdt\Big]<\infty$ and $E\Big[\int_{0}^{T}\widehat{\pi}^2_t\mathcal{V}_tdt\Big]<\infty$, and there exists a unique strong solution to \eqref{wealth process}. The set of all admissible strategies is denoted by $\widehat{\Pi}$.

The aim of the insurer is to find the best strategy to maximize the expected utility of the terminal wealth with minimum guaranteed threshold $L\geq 0$, that is
\begin{equation}\label{utility maximization}
\sup_{\widehat{u}\in{\widehat{\Pi}}}E\big[U(\widehat{X}_T-L)\big],
\end{equation}
where $\widehat{X}_t$ is subject to \eqref{wealth process} and $\widehat{X}_T\geq{L}$.
To solve utility maximization \eqref{utility maximization}, using
$D(t):=c(\eta-\vartheta)\int^{T}_{t}e^{-r(s-t)}ds$, $X_t:=\widehat{X}_t+D(t)$, $\pi_t:=\widehat{\pi}_t/X_t$ and $q_t:=\widehat{q}_t/X_t$, the wealth process \eqref{wealth process} is rewritten as
\begin{equation}\label{convert wealth process}
dX_t=X_t\Big[rdt+\lambda\pi_t\mathcal{V}_tdt +\pi_t\sqrt{\mathcal{V}_t}dW_{t}^{1}+c\vartheta{q}_tdt+b{q}_tdW_t\Big].
\end{equation}
Obviously, the reinsurance-investment strategy $(q_t,\pi_t)$ in \eqref{convert wealth process} is also the admissible strategy and its admissible set is denoted by $\Pi$. Note that $X_T=\widehat{X}_T$. Thus, the equivalent form of utility maximization problem \eqref{utility maximization} is
\begin{equation}\label{convert utility maximization}
\sup_{(q,\pi)\in{{\Pi}}}E\big[U({X}_T-L)\big],
\end{equation}
For the Heston model, we define the value function as
\begin{equation}\label{value-function-heston}
V(t,x,v) = \sup_{(q,\pi)\in{{\Pi}}}E\big[U\big(X_T-L\big)|X_t=x,\mathcal{V}_t=v\big],\notag
\end{equation}
where $X_t$ satisfies \eqref{convert wealth process}, $X_T\geq L$ and $x=\widehat{x}+c(\eta-\vartheta){\displaystyle \int^{T}_{t}}e^{-rs}ds$.
The terminal condition is
\begin{equation}\label{primal terminal condition}
V(T,x,v)=U\big(x-L\big).\notag
\end{equation}
Now we apply Algorithm~\ref{alg:multi-D} to solve the problem.
Denote the infinitesimal generator of \eqref{convert wealth process} by ${\mathcal L}^{q,\pi}$. Then we derive that
\begin{align}
&{\mathcal L}^{q,\pi}\left(g_{k_{1}}^{(1)}(\tilde{x}(x)) g_{k_{2}}^{(2)}(\tilde{v}(v))\right)\notag\\
&= (\lambda\pi v+c\vartheta q)x\frac{d g_{k_{1}}^{(1)}(\tilde{x}(x)) }{d x} g_{k_{2}}^{(2)}(\tilde{v}(v))+\kappa(\theta-v) g_{k_{1}}^{(1)}(\tilde{x}(x))\frac{d g_{k_{2}}^{(2)}(\tilde{v}(v))}{d v}\notag\\
& \quad +\frac{\sigma^2 v}{2}g_{k_{1}}^{(1)}(\tilde{x}(x)) \frac{d^2 g_{k_{2}}^{(2)}(\tilde{v}(v))}{d v^2} +\frac{(\pi^2 v+b^2q^2)x^2}{2}  \frac{d^2 g_{k_{1}}^{(1)}(\tilde{x}(x))}{d x^2}g_{k_{2}}^{(2)}(\tilde{v}(v))\notag\\
& \quad + \rho \pi \sigma x v  \frac{d g_{k_{1}}^{(1)}(\tilde{x}(x)) }{d x}\frac{d g_{k_{2}}^{(2)}(\tilde{v}(v))}{d v},\notag
\end{align}
where $\hat{x},\, \hat{v}, \,\tilde{x},\,\tilde{v}$ are given by \eqref{hat_x} - \eqref{tilde_v}.
The strategies $\pi^{\ast}$ is given by \eqref{strategy-Heston-1} and $q^{\ast}$ is given by
\begin{align}\label{strategy-reinsurance-Heston-1}
q^{*}=-
\frac{c\vartheta\sum\limits_{k_{1}, k_2=0}^{\infty }
\left( \int_{\Omega^{2}} V(t+h, \hat{x}(y_1),\hat{v}(y_2)) g_{k_{1}}^{(1)}(y_{1}) g_{k_{2}}^{(2)}(y_{2}) d{\bf y}\right)
 \frac{d g_{k_{1}}^{(1)}(\tilde{x}(x))}{d x}g_{k_{2}}^{(2)}(\tilde{v}(v))}
 {b^2x\sum\limits_{k_{1}, k_2=0}^{\infty }
\left( \int_{\Omega^{2}} V(t+h, \hat{x}(y_1),\hat{v}(y_2)) g_{k_{1}}^{(1)}(y_{1}) g_{k_{2}}^{(2)}(y_{2}) d{\bf y}\right)\frac{d^{2}g_{k_{1}}^{(1)}(\tilde{x}(x))}{d x^{2}} g_{k_{2}}^{(2)}(\tilde{v}(v))}.\notag
\end{align}
Then the rest steps are similar to Algorithm~\ref{alg:2-D}.

For rough Heston model, the fractional kernel can be expressed as the Laplace transform of a positive measure $\zeta$
\[
K(t)=\int^\infty_0 e^{-\gamma t}\zeta(d\gamma),\quad \zeta(d\gamma)=\frac{\gamma^{-\alpha}}{\Gamma(\alpha)\Gamma(1-\alpha)}d\gamma.
\]
Then $\zeta$ is approximated by a finite sum of Dirac measures $\zeta^{\ell}=\sum^{\ell}_ic^{\ell}_i\delta_{\gamma^{\ell}_i}$ with positive weights $(c^{\ell}_i)_{1\leq i\leq {\ell}}$ and mean reversions $(\gamma^{\ell}_i)_{1\leq i\leq {\ell}}$ for ${\ell}\geq 1$. This in turn yields an approximation
of the fractional kernel by a sequence of smoothed kernels $(K^{\ell})_{{\ell}\geq 1}$ given by
\[
K^{\ell}(t)=\sum^{\ell}_{i=1}c^{\ell}_ie^{-\gamma^{\ell}_i t},\quad {\ell}\geq 1.
\]
Therefore, the stock price process can be well approximated by the multi-factor stochastic volatility model (see \cite{Jaber2019a}, \cite{Jaber2019c}) as follows
\begin{equation}\label{stock process model}
\frac{dS^{\ell}_t}{S^{\ell}_t}=(r+\lambda\mathcal{V}^{\ell}_t)dt+\sqrt{\mathcal{V}^{\ell}_t}dW_t^{1},\notag
\end{equation}
and the variance process $\mathcal{V}^{\ell}_t$ is given by a fractional
square-root process
\begin{equation}\label{variance process}
\mathcal{V}^{\ell}_t=\xi(t)+\sum^{{\ell}}_{i=1}c^{\ell}_i\mathcal{V}^{{\ell},i}_t,\notag
\end{equation}
and
\begin{equation}\label{variance process}
d\mathcal{V}^{{\ell},i}_t=(-\gamma^{\ell}_i\mathcal{V}^{{\ell},i}_t-\kappa \mathcal{V}^{\ell}_t)dt+\sigma\sqrt{\mathcal{V}^{\ell}_t}dW_t^{2},\notag
\end{equation}
where $\xi(t)=\mathcal{V}_0+\kappa\theta{\displaystyle \int^t_0K(t-s)ds}$. These inspire us to approximate the transformed wealth process as follows
\begin{equation}\label{approximated wealth process}
dX^{\ell}_t=X^{\ell}_t\Big[(r+\lambda\pi_t\mathcal{V}^{\ell}_t)dt +\pi_t\sqrt{\mathcal{V}^{\ell}_t}dW_t^{1}+c\vartheta{q}_tdt+b{q}_tdW_t\Big].
\end{equation}
The utility maximization problem \eqref{convert utility maximization} is approximated by
\begin{equation}\label{approximated utility maxization}
\sup_{(q,\pi)\in\Pi^{\ell}}{\Bbb E}\left[U\big(X^{\ell}_T-L\big)\right],
\end{equation}
where $X^{\ell}_t$ is subject to \eqref{approximated wealth process}. Therefore, the utility maximization problem \eqref{approximated utility maxization} can be solved by principle of dynamic programming.
Define the value function as follows
\begin{equation}\label{approximated value function}
V(t,x,v_1,\ldots,v_{\ell})=\sup_{(q,\pi)\in\Pi^{\ell}}{\Bbb E}\left[U\big(X^{\ell}_T-L\big)|X_t=x,\mathcal{V}^{\ell,1}_t=v_1,\ldots,\mathcal{V}^{\ell,\ell}_t=v_\ell\right],\notag
\end{equation}
with terminal condition
\begin{equation}\label{boundary-condition-rough}
V(T,x,v_1,\ldots,v_{\ell})=U(x-L).\notag
\end{equation}
Then the infinitesimal generator of \eqref{approximated wealth process} is given by
\begin{align}
&{\mathcal L}^{q,\pi} f(x,v_1,\ldots,v_{\ell})\notag\\
 &=\left[r+\lambda\pi\left(\xi(t)+\sum^{\ell}_{l=1}c^{\ell}_lv_l\right)+c\vartheta q\right]xf_{x}
 + \frac{1}{2}x^2\left[\pi^2\left(\xi(t)+\sum^{\ell}_{l=1}c^{\ell}_lv_l\right)+b^2q^2\right]f_{xx}\notag\\
 &-\sum^{\ell}_{i=1} \left(\gamma^{\ell}_iv_i+\kappa \xi(t)+\kappa\sum^{\ell}_{l=1}c^{\ell}_lv_l\right)f_{v_i}
+\,\frac{1}{2}\sum^{\ell}_{i,j=1}\sigma^2\left(\xi(t)+\sum^{\ell}_{l=1}c^n_lv_l\right)f_{v_iv_j} \notag\\
& +\sum^{\ell}_{i=1}\rho\sigma x\pi \left(\xi(t)+\sum^{\ell}_{l=1}c^{\ell}_lv_l\right)f_{xv_i}.\notag
\end{align}
For ease of exposition, we denote $v_0:=x$ and
${\bf v}:=(v_0, v_1,\ldots,v_{\ell})$ and for the computation we set $v_l\in[v_{\min}^l,v_{\max}^l]\, l=0,1,\ldots, \ell$. Using the following transformation
\begin{align}
\hat{\mathbf{v}}(\mathbf{y}) &= \left(\hat{v}_0(y_0),\hat{v}_1(y_1),\ldots,\hat{v}_\ell(y_\ell)\right),\notag\\
\tilde{\mathbf{v}}(\mathbf{v}) &= \left(\tilde{v}_0(v_0),\tilde{v}_1(v_1),\ldots,\tilde{v}_\ell(v_\ell)\right),\notag\\
\hat{v}_l(y_l) &= (v_{\max}^l-v_{\min}^l)(y_l+1)/2+v_{\min}^l,\notag\\
\tilde{v}_l(v_l) &= \frac{2(v_l-v_{\min}^l)}{v_{\max}^l-v_{\min}^l}-1,\notag
\end{align}
where $\mathbf{y}=(y_0,y_1,\ldots,y_{\ell})\in \Omega^{\ell+1}:=[-1,1]^{\ell+1}$, similarly to the derivation in Section~\ref{sec:multi-D}, we obtain that
\begin{align}\label{multi-app-exam}
V(t, {\bf v})\approx\sup\limits_{(q,\pi)\in\Pi} \sum\limits_{k_{0}, k_{1}, \ldots, k_{\ell}=0}^{\infty }&\left( \int_{\Omega^{\ell+1}} V(t+h, \hat{\mathbf{v}}(\mathbf{y}))  \prod\limits_{i=0}^{\ell}g_{k_{i}}^{(i)}(y_{i})d{\bf y}\right) \notag\\
&\cdot\left( \prod\limits_{i=0}^{\ell}g_{k_{i}}^{(i)}(\tilde{v}_i(v_i)) + h {\mathcal L}^{q,\pi}\left(\prod\limits_{i=0}^{\ell}g_{k_{i}}^{(i)}(\tilde{v}_i(v_i))\right) \right).\notag
\end{align}
Denote
\[
V_{{\bf k}}(t):=\int_{\Omega^{\ell+1}} V(t, \hat{\mathbf{v}}(\mathbf{y})) \prod\limits_{i=0}^{\ell}g_{k_{i}}^{(i)}(y_{i})d{\bf y},\quad k_j=0,1,\ldots;\, j=0,1,\ldots,\ell.
\]
The first order condition gives that
\begin{align}\label{strategy-multi-D-exam}
\pi^*=
&-\frac{\sum\limits_{k_{0}, k_{1}, \ldots, k_{\ell}=0}^{\infty }V_{{\bf k}}(t)\left(\lambda\left(\frac{d g_{k_0}^{(0)}(\tilde{v}_0(x))}{dx} \prod\limits_{j=1}^{\ell}g_{k_{j}}^{(j)}(\tilde{v}_j(v_j))\right)\right)}
{x \sum\limits_{k_{0}, k_{1}, \ldots, k_{\ell}=0}^{\infty }V_{{\bf k}}(t)
\frac{d^2 g_{k_0}^{(0)}(\tilde{v}_0(x))}{dx^2}\prod\limits_{i=1}^{\ell}g_{k_{i}}^{(i)}(\tilde{v}_i(v_i))}\notag\\
&-\frac{\sum\limits_{k_{0}, k_{1}, \ldots, k_{\ell}=0}^{\infty }V_{{\bf k}}(t)\left(\rho \sigma  \sum\limits_{i=1}^{\ell}\left(\frac{d g_{k_0}^{(0)}(\tilde{v}_0(x))}{dx} \frac{d g_{k_{i}}^{(i)}(\tilde{v}_i(v_i))}{dv_i} \prod\limits_{j=1,j\neq i}^{\ell}g_{k_{j}}^{(j)}(\tilde{v}_j(v_j))\right)\right)}
{x \sum\limits_{k_{0}, k_{1}, \ldots, k_{\ell}=0}^{\infty }V_{{\bf k}}(t)
\frac{d^2 g_{k_0}^{(0)}(\tilde{v}_0(x))}{dx^2}\prod\limits_{i=1}^{\ell}g_{k_{i}}^{(i)}(\tilde{v}_i(v_i))},\notag\\
q^* = &-\frac{c\vartheta\sum\limits_{k_{0}, k_{1}, \ldots, k_{\ell}=0}^{\infty }V_{{\bf k}}(t)\left(\frac{d g_{k_0}^{(0)}(\tilde{v}_0(x))}{dx} \prod\limits_{j=1}^{\ell}g_{k_{j}}^{(j)}(\tilde{v}_j(v_j))\right)}
{b^2 x \sum\limits_{k_{0}, k_{1}, \ldots, k_{\ell}=0}^{\infty }V_{{\bf k}}(t)
\frac{d^2 g_{k_0}^{(0)}(\tilde{v}_0(x))}{dx^2}\prod\limits_{i=1}^{\ell}g_{k_{i}}^{(i)}(\tilde{v}_i(v_i))}\vee 0.\notag
\end{align}
Then Algorithm~\ref{alg:multi-D} is reduced to  the following specialized Algorithm \ref{alg:multi-D-exam} for the optimal reinsurance-investment under the rough Heston model.

\begin{exam}\label{exam:Reinsurance}
In this example, we simulate the paths of wealth process, volatility process and optimal investment and reinsurance strategies under Heston model and rough Heston model with power utility in Example~\ref{exam:HestonPower} by using the method of delta family approach. For the reinsurance process, the model parameters are taken from \cite{Bi and Cai2019}
\begin{equation}
c=0.13,\, b=0.6,\, \eta=0.3,\, \vartheta=0.5.\notag
\end{equation}
The minimum guaranteed threshold $L=0$. For Heston model, The parameters are taken the same as Example \ref{exam:HestonPower}. For rough Heston model, we use multi-factor stochastic volatility model to approximate it. We set the number of factors $\ell=3$ and the value of model parameters are taken from \cite{Abi2019}
\begin{equation}
\mathcal{V}_0=0.02,\, \ell=3,\, \lambda=0.5,\, \rho=-0.7,\, \kappa=0.3,\, \theta=0.02,\, \sigma=0.3.\notag
\end{equation}
The interest rate is $r=0.05$, the initial value of wealth $\widehat{x}_0=5$ and the investment horizon $T=1$.
The initial value of converted wealth is calculated as $x=\widehat{x}_0+c(\eta-\vartheta)\int^{T}_{0}e^{-rs}ds=4.975$.
\end{exam}

\begin{algorithm}[H]
\caption{(Optimal reinsurance-investment under rough Heston models)}\label{alg:multi-D-exam}
\begin{algorithmic}[1]
\STATE  Denote $V^{n}_{{\bf m}}\approx V_{{\bf m}}(t_{n})$ for ${\bf m}=(m_0, m_1,\ldots,m_{\ell})$;
$m_j=0,1,\ldots, M_j$; $j=0,1,\ldots, \ell$; $n=0,1,\ldots, N$.
\STATE  The algorithm is given by the following recursion for $n=0,1,\ldots,N-1$,
\begin{align}\label{value-multi-D-algorithm-exam}
V_{{\bf m}}^{n}&\approx V_{{\bf m}}^{n+1} + h\sum\limits_{k_{0}=0}^{M_0 }\cdots\sum\limits_{k_{\ell}=0}^{M_{\ell}}
 V_{{\bf k}}^{n+1} \int_{{\Omega}^{\ell+1}}\prod\limits_{j=0}^{\ell}g_{m_{j}}^{(j)}(y_j) {\mathcal L}^{\pi^{*}_{n+1}}\left(\prod\limits_{i=0}^{\ell}g_{k_{i}}^{(i)}(y_i)\right)d{\bf v},
\end{align}
with $\pi^{\ast}_{n+1}$ and $q^{\ast}_{n+1}$ given by

\begin{align}
\pi^{\ast}_{n+1}=
&-\frac{\sum\limits_{k_{0}, k_{1}, \ldots, k_{\ell}=0}^{\infty }V_{{\bf k}}^{n+1}\left(\lambda\left(\frac{d g_{k_0}^{(0)}(\tilde{v}_0(x))}{dx} \prod\limits_{j=1}^{\ell}g_{k_{j}}^{(j)}(\tilde{v}_j(v_j))\right)\right)}
{x \sum\limits_{k_{0}, k_{1}, \ldots, k_{\ell}=0}^{\infty }V_{{\bf k}}^{n+1}
\frac{d^2 g_{k_0}^{(0)}(\tilde{v}_0(x))}{dx^2}\prod\limits_{i=1}^{\ell}g_{k_{i}}^{(i)}(\tilde{v}_i(v_i))}\notag\\
&-\frac{\sum\limits_{k_{0}, k_{1}, \ldots, k_{\ell}=0}^{\infty }V_{{\bf k}}^{n+1}\left(\rho \sigma  \sum\limits_{i=1}^{\ell}\left(\frac{d g_{k_0}^{(0)}(\tilde{v}_0(x))}{dx} \frac{d g_{k_{i}}^{(i)}(\tilde{v}_i(v_i))}{dv_i} \prod\limits_{j=1,j\neq i}^{\ell}g_{k_{j}}^{(j)}(\tilde{v}_j(v_j))\right)\right)}
{x \sum\limits_{k_{0}, k_{1}, \ldots, k_{\ell}=0}^{\infty }V_{{\bf k}}^{n+1}
\frac{d^2 g_{k_0}^{(0)}(\tilde{v}_0(x))}{dx^2}\prod\limits_{i=1}^{\ell}g_{k_{i}}^{(i)}(\tilde{v}_i(v_i))}\notag\\
q^{\ast}_{n+1} = &-\frac{c\vartheta\sum\limits_{k_{0}, k_{1}, \ldots, k_{\ell}=0}^{\infty }V_{{\bf k}}^{n+1}\left(\frac{d g_{k_0}^{(0)}(\tilde{v}_0(x))}{dx} \prod\limits_{j=1}^{\ell}g_{k_{j}}^{(j)}(\tilde{v}_j(v_j))\right)}
{b^2 x \sum\limits_{k_{0}, k_{1}, \ldots, k_{\ell}=0}^{\infty }V_{{\bf k}}^{n+1}
\frac{d^2 g_{k_0}^{(0)}(\tilde{v}_0(x))}{dx^2}\prod\limits_{i=1}^{\ell}g_{k_{i}}^{(i)}(\tilde{v}_i(v_i))}\vee 0.\notag
\end{align}
The terminal condition
\begin{equation}\label{terminal-condition-multi-D-exam}
V^{N}_{{\bf m}}=\int_{\Omega^{\ell+1}}U(\hat{v}(y_0))\prod\limits_{j=0}^{\ell}g_{m_{j}}^{(j)}(y_j)d {\bf y}.
\end{equation}
The integrals in \eqref{value-multi-D-algorithm-exam} and \eqref{terminal-condition-multi-D-exam} are evaluated by Gauss-Legendre quadrature rules.
\STATE Finally we obtain the approximation of the value functions
\begin{equation}\label{approx-value-function-multi-D-exam}
 V(t_{n},{\bf v})\approx \sum\limits_{m_{0}=0}^{M_0 }\cdots\sum\limits_{m_{\ell}=0}^{M_{\ell}} V_{{\bf m}}^{n}\prod\limits_{j=0}^{\ell}g_{m_{j}}^{(j)}(v_j),\notag
\end{equation}
and the optimal strategies
\begin{align}\label{strategy-approx-multi-D-exam}
\pi^{*}(t_{n},{\bf v})&\approx \pi^{*}_{n}({\bf v}),\quad n=0,1,\ldots,N,\notag
\end{align}
where $v_0=x$.
\end{algorithmic}
\end{algorithm}

\begin{figure}[!htbp]
\centering
\subfigure{
\begin{minipage}[t]{0.5\linewidth}
\centering
\includegraphics[height=20cm,width=8cm]{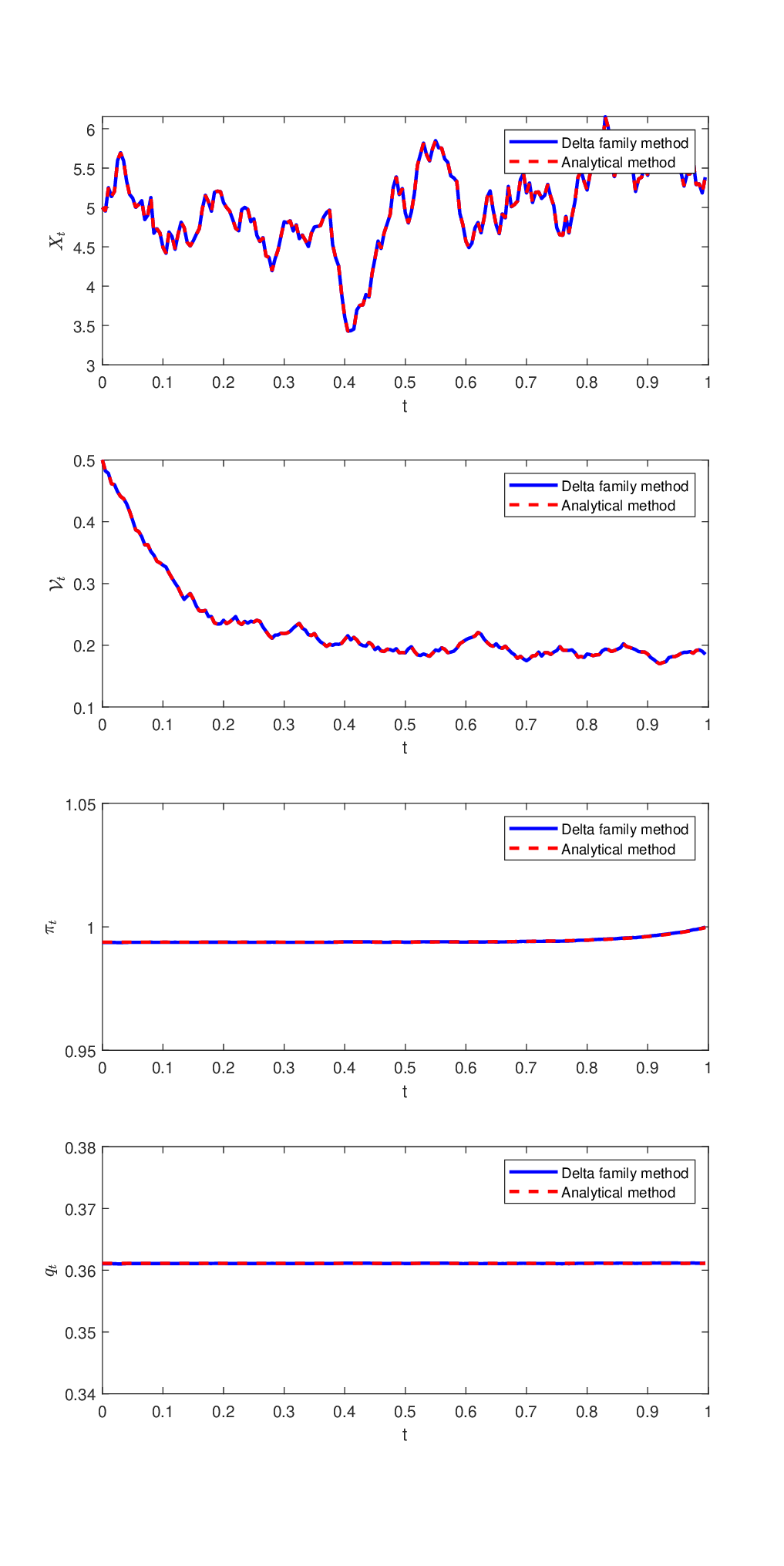}
\end{minipage}%
}%
\subfigure{
\begin{minipage}[t]{0.5\linewidth}
\centering
\includegraphics[height=20cm,width=8cm]{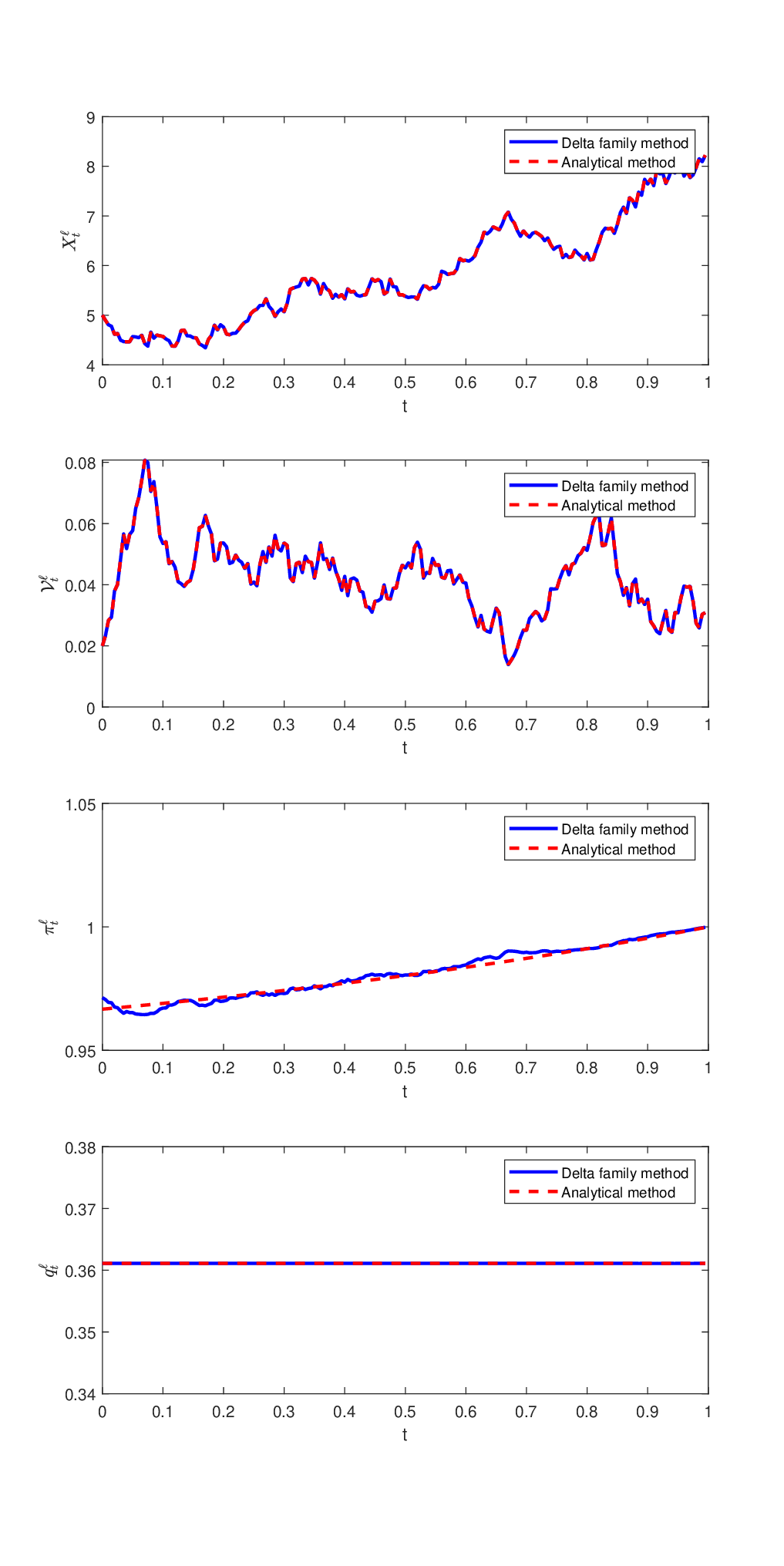}
\end{minipage}%
}
\caption{The paths of wealth and optimal reinsurance-investment strategies under Heston model (left) and rough Heston model (right) for Example \ref{exam:Reinsurance}. }
\label{Figure:Reinsurance-Path}
\end{figure}

For the computation of delta family approach, we set
\[
M=20,\,N=2000,\,x_{\min}=1,\,x_{\max}=20,
\]
and for Heston model
\[
v_{\min}=0,\,v_{\max}=1,
\]
and for multi-factor stochastic volatility model
\[
v^l_{\min}=-0.2,\,v^l_{\max}=0.2,
~~\text{for}~~l=1,2,\ldots,\ell.
\]
The paths of wealth process and the strategies are drawn using the delta family methods and
the analytical methods in \cite{Ma2020} or \cite{Kraft2005} for Heston model and \cite{Ma2022} for multi-factor stochastic volatility model. From Figure~\ref{Figure:Reinsurance-Path}, we observe that the delta family methods are pretty accurate.

\subsection{Optimal control stopping problem under SLV models}

In this section, we solve the optimal control stopping problems under SLV models using the delta family methods combined with operator splitting technique.
Consider a fixed time horizon $[0,T]$. Let the stochastic processes and random variables be defined on a filtered probability space $\left(\Omega,F,\{\mathcal{F}_t\}_{0\leq{t}\leq{T}},P\right)$, where $P$ is the probability measure, and $F=\{\mathcal{F}_t\}_{0\leq{t}\leq{T}}$ is the filtration generated
by two standard Brownian motions $W_t^{1}$ and $W_t^{2}$ with $d\langle W_t^{1}, W_{t}^{2}\rangle=\rho{dt}$ and $-1\leq\rho\leq 1$. In addition, we assume that the markets do not have transaction costs or taxes, and the trading happens continuously.

We assume that the financial market consists of one risk-free asset with interest rate zero and one risky asset. The price of the risky asset $S_t$ is subject to the SLV model as follows
\begin{equation}\label{SLV stock process model}
\frac{dS_t}{S_t}=\omega(S_t,\mathcal{V}_t)dt+\eta(\mathcal{V}_t)\Gamma(S_t)dW_t^{1},\notag
\end{equation}
and the variance process $\mathcal{V}_t$ is given by a fractional
square-root process
\begin{equation}\label{SLV variance process}
d\mathcal{V}_t=\beta(\mathcal{V}_t)dt+\zeta(\mathcal{V}_t)dW_s^{2}.\notag
\end{equation}
\begin{remark}
Some typical SLV models are listed as follows:
\begin{itemize}
\item[(i)] Heston model
\begin{align}\label{SLV-Heston}
&\omega(S_t,\mathcal{V}_t)=(r+\lambda\mathcal{V}_t)S_t,~~\eta(\mathcal{V}_t)=\sqrt{\mathcal{V}_t},~~\Gamma(S_t)=S_t,\notag\\
&\beta(\mathcal{V}_t)=\kappa(\theta-\mathcal{V}_t),~~\zeta(\mathcal{V}_t)=\sigma\sqrt{\mathcal{V}_t}.
\end{align}

\item[(ii)] 4/2 model
\begin{align}\label{SLV-42}
&\omega(S_t,\mathcal{V}_t)=(r+\lambda\mathcal{V}_t)S_t,~~\eta(\mathcal{V}_t)=a\sqrt{\mathcal{V}_t}+b/\sqrt{\mathcal{V}_t},~~\Gamma(S_t)=S_t,\notag\\
&\beta(\mathcal{V}_t)=\kappa(\theta-\mathcal{V}_t),~~\zeta(\mathcal{V}_t)=\sigma\sqrt{\mathcal{V}_t}.
\end{align}

\item[(iii)] $\alpha$-Hyper model
\begin{align}\label{SLV-alpha}
&\omega(S_t,\mathcal{V}_t)=(r+\lambda\mathcal{V}_t)S_t, ~~\eta(\mathcal{V}_t)=\exp({\mathcal{V}_t}),~~\Gamma(S_t)=S_t,\notag\\
&\beta(\mathcal{V}_t)=\kappa(\theta-\exp(a\mathcal{V}_t)), ~~\zeta(\mathcal{V}_t)=\sigma\sqrt{\mathcal{V}_t}.
\end{align}
\end{itemize}
\end{remark}
We study the following problem with its value function characterized by
\begin{equation}
V(t,x,s,v) = \sup_{\tau\in{[t,T]},\,\pi\in\mathcal{A}}{\Bbb E}[e^{-\gamma(\tau-t)}U(X_{\tau}-L)\mid X_t=x, \mathcal{V}_t=v, S_t=s],\notag
\end{equation}
where $L>0$ is the minimum wealth threshold value and $\gamma$ is utility discount factor.
From \cite{Guan2017}, the value function $V$ satisfies the following HJB variational inequality
\begin{equation}\label{SLV-HJB}
\min\Big\{-\frac{\partial V}{\partial t}-\sup_{\pi\in\mathcal{A}}\mathcal{L}^{\pi}[V],\, V-G(x)\Big\},
\end{equation}
where $G(x)=U(x-L)$ and
\begin{align}
\mathcal{L}^{\pi}[f(t,x,s,v)] = &-\gamma f +\Big[r+\Big(\frac{\omega(s,v)}{s}-r\Big)\pi\Big]xf_x+ \frac{1}{2}\frac{\eta^2(v)\Gamma^2(s)}{s^2}\pi^2x^2 f_{xx}\notag\\
&+\omega(s,v)f_s+\frac{1}{2}\eta^2(v)\Gamma^2(s)f_{ss}+\beta(v)f_v+\frac{1}{2}\zeta(v)f_{vv}\notag\\
&+\eta^2(v)\Gamma^2(s)\frac{\pi x}{s}f_{xs}+\zeta(v)\eta(v)\Gamma(s)\rho \Big(\frac{\pi x}{s}f_{xv}+f_{sv}\Big),\notag
\end{align}
with terminal condition
\[
V(T,x,s,v)=G(x),\quad x\in[0,\infty).
\]
The HJB variational inequality \eqref{SLV-HJB} can be transformed as a nonlinear complementarity problem
\begin{align}
&-V_t\geq \sup_{\pi\in\mathcal{A}}\mathcal{L}^{\pi}[V],\label{NLCP1}\\
&V\geq G,\label{NLCP2}\\
&(V_t-\mathcal{L}V)(V-G)=0.\label{NLCP3}
\end{align}
Adding the slack function $\psi(t,x,s,v)$, we rewrite \eqref{NLCP1} - \eqref{NLCP3} as
\begin{align}
&-V_t= \sup_{\pi\in\mathcal{A}}\mathcal{L}^{\pi}[V]+\psi,\label{NLCP4}\\
&V\geq G,\\
&\psi\geq 0,\\
&\psi(V-G)=0.\label{NLCP7}
\end{align}
Let $V^n(x,s,v)\approx V(t_n,x,s,v)$ and $\psi^n(x,s,v)\approx \psi(t_n,x,s,v)$. Then
Using the operator splitting technique (see \cite{Ikonen2004} and \cite{Chen2020}), \eqref{NLCP4} - \eqref{NLCP7} are discretized in time by
\begin{align}
&\widetilde{V}^n=V^{n+1}+\sup_{\pi\in\mathcal{A}}h\Big(\mathcal{L}^{\pi}[V^{n+1}]+\psi^{n+1}\Big),\label{split-1}\\
&\frac{V^{n}-\widetilde{V}^n}{h}=\psi^{n}-\psi^{n+1},\label{split-2}\\
&\psi^{n}\geq0,\quad V^n\geq0,\quad \psi^{n}(V^n-G)=0.\label{split-3}
\end{align}
Furthermore \eqref{split-2} and \eqref{split-3} give explicitly that
\begin{equation}\label{solution-splits-2-3}
(\psi^{n}, V^n)=\left\{
 \begin{array}{ll}
\Big(0,\widetilde{V}^n-h\psi^{n+1}\Big),& \text{if}\quad \widetilde{V}^n-h\psi^{n+1}\geq G,\\
\Big(\frac{G-\widetilde{V}^n}{h}+\psi^{n+1},G\Big),& \text{otherwise}.
 \end{array}
 \right.
\end{equation}
Now we derive the delta family methods to discretize \eqref{split-1} and \eqref{solution-splits-2-3} in space.  Denote $v_0=x$, $v_1=s$, $v_2=v$ and
\begin{align*}
 {\bf v}&:=(v_0, v_1,v_2),
\end{align*}
We set $v_l\in[v_{\min}^l,v_{\max}^l]$, $l=0,1,2$ and use the following transformation
\begin{align}
\hat{\mathbf{v}}(\mathbf{y}) &= \left(\hat{v}_0(y_0),\hat{v}_1(y_1),\hat{v}_2(y_2)\right),\notag\\
\tilde{\mathbf{v}}(\mathbf{v}) &= \left(\tilde{v}_0(v_0),\tilde{v}_1(v_1),\tilde{v}_2(v_2)\right),\notag\\
\hat{v}_l(y_l) &= (v_{\max}^l-v_{\min}^l)(y_l+1)/2+v_{\min}^l,\notag\\
\tilde{v}_l(v_l) &= \frac{2(v_l-v_{\min}^l)}{v_{\max}^l-v_{\min}^l}-1,\notag
\end{align}
where $\mathbf{y}=(y_0,y_1,y_2)\in \Omega^3:=[-1,1]^{3}$, to give that
\begin{align}
\widetilde{V}^n({\bf v})&\approx \sum\limits_{k_{0}, k_{1},k_{2}=0}^{\infty } \left( \int_{\Omega^{3}} V^{n+1}(\hat{\mathbf{v}}(\mathbf{y}))  \prod\limits_{i=0}^{2}g_{k_{i}}^{(i)}(y_{i})d{\bf y}\right) \notag\\
&\cdot\left( (1-\gamma h)\prod\limits_{i=0}^{2}g_{k_{i}}^{(i)}(\tilde{v}_i(v_i)) + h\cdot {\mathcal L}^{\pi^*_{n+1}}\left(\prod\limits_{i=0}^{2}g_{k_{i}}^{(i)}(\tilde{v}_i(v_i))\right) \right)\notag\\
&+\sum\limits_{k_{0}, k_{1},k_{2}=0}^{\infty }\left( \int_{\Omega^{3}} \psi^{n+1}( \hat{\mathbf{v}}(\mathbf{y}))  \prod\limits_{i=0}^{2}g_{k_{i}}^{(i)}(y_{i})d{\bf y}\prod\limits_{i=0}^{2}g_{k_{i}}^{(i)}(\tilde{v}_i(v_i))\right).\notag
\end{align}
Denote
\begin{align*}
V^n_{{\bf k}}:=\int_{\Omega^{3}} V^n(\hat{\mathbf{v}}(\mathbf{y}))  \prod\limits_{i=0}^{2}g_{k_{i}}^{(i)}(y_{i})d{\bf y},\\
\psi^n_{\bf k}:=\int_{\Omega^{3}} \psi^n(\hat{\mathbf{v}}(\mathbf{y}))  \prod\limits_{i=0}^{2}g_{k_{i}}^{(i)}(y_{i})d{\bf y},\\
\widetilde{V}^n_{{\bf k}}:=\int_{\Omega^{3}} \widetilde{V}^n(\hat{\mathbf{v}}(\mathbf{y}))  \prod\limits_{i=0}^{2}g_{k_{i}}^{(i)}(y_{i})d{\bf y},
\end{align*}
for ${\bf k}=(k_1,\, k_2,\, k_3)$; $k_i=0,1,\ldots, M_i$; $i=1,2,3$; $n=0,1,\ldots, N$.
Then $\pi^*_{n}({\bf v})$ is given by
\begin{align}\label{SLV-optimal-strategy}
\pi^*_{n}({\bf v})= &
\frac{(\omega(v_1,v_2)-rv_1)v_1\sum\limits_{k_{0}=0}^{M_0 }\sum\limits_{k_{1}=0}^{M_{1}}\sum\limits_{k_{2}=0}^{M_{2}}V^n_{{\bf k}}\left(\frac{d g_{k_0}^{(0)}(\tilde{v}_0(v_0))}{dv_0} g_{k_{1}}^{(1)}(\tilde{v}_1(v_1))g_{k_{2}}^{(2)}(\tilde{v}_2(v_2))\right)}{\eta^2(v_2)\Gamma^2(v_1)v_0 \sum\limits_{k_{0}=0}^{M_0 }\sum\limits_{k_{1}=0}^{M_{1}}\sum\limits_{k_{2}=0}^{M_{2}}V^n_{{\bf k}}\left(\frac{d^2 g_{k_0}^{(0)}(\tilde{v}_0(v_0))}{dv_0^2}g_{k_{1}}^{(1)}(\tilde{v}_1(v_1))g_{k_{2}}^{(2)}(\tilde{v}_2(v_2))\right)}\notag\\
&+\frac{\zeta(v_2)\rho v_1\sum\limits_{k_{0}=0}^{M_0 }\sum\limits_{k_{1}=0}^{M_{1}}\sum\limits_{k_{2}=0}^{M_{2}}V^n_{{\bf k}}\left(\frac{d g_{k_0}^{(0)}(\tilde{v}_0(v_0))}{dv_0} \frac{d g_{k_1}^{(1)}(\tilde{v}_1(v_1))}{dv_1}g_{k_{2}}^{(2)}(\tilde{v}_2(v_2))\right)}{\eta(v_2)\Gamma(v_1)v_0 \sum\limits_{k_{0}=0}^{M_0 }\sum\limits_{k_{1}=0}^{M_{1}}\sum\limits_{k_{2}=0}^{M_{2}}V^n_{{\bf k}}\left(\frac{d^2 g_{k_0}^{(0)}(\tilde{v}_0(v_0))}{dv_0^2}g_{k_{1}}^{(1)}(\tilde{v}_1(v_1))g_{k_{2}}^{(2)}(\tilde{v}_2(v_2))\right)}\notag\\
&+\frac{v_1\sum\limits_{k_{0}=0}^{M_0 }\sum\limits_{k_{1}=0}^{M_{1}}\sum\limits_{k_{2}=0}^{M_{2}}V^n_{{\bf k}}\left(\frac{d g_{k_0}^{(0)}(\tilde{v}_0(v_0))}{dv_0} g_{k_{1}}^{(1)}(\tilde{v}_1(v_1))\frac{d g_{k_2}^{(2)}(\tilde{v}_2(v_2))}{dv_2}\right)}{v_0 \sum\limits_{k_{0}=0}^{M_0 }\sum\limits_{k_{1}=0}^{M_{1}}\sum\limits_{k_{2}=0}^{M_{2}}V^n_{{\bf k}}\left(\frac{d^2 g_{k_0}^{(0)}(\tilde{v}_0(v_0))}{dv_0^2}g_{k_{1}}^{(1)}(\tilde{v}_1(v_1))g_{k_{2}}^{(2)}(\tilde{v}_2(v_2))\right)}.
\end{align}
Therefore Algorithm~\ref{alg:multi-D} is  reduced to the following specialized Algorithm \ref{alg:SLV-D} for the optimal control stopping problem under SLV models.

\begin{exam}\label{exam:SLVStopping}
In this example, we solve the optimal stopping investment problem with power utility.
For Heston model \eqref{SLV-Heston}, The values of model parameters are taken as same as that in Example~\ref{exam:HestonPower}, the utility discount factor $\gamma=0.15$ and $L=1.0$. For 4/2 stochastic volatility model \eqref{SLV-42}, the values of model parameters are taken from \cite{Grasselli2017}
\begin{eqnarray}
\rho=-0.7,\,\sigma=0.2,\,\kappa=1.8,\,\theta=0.04,\,r=0.02,\,T=1,\,a=0.5,\,b=0.04.
\end{eqnarray}
In addition, the utility discount factor $\gamma=0.05$ and $L=1.0$. For $\alpha$-Hypergeometric stochastic volatility model \eqref{SLV-alpha}, we set $a=0.01$ and the other parameters are taken as same as 4/2 model.
\end{exam}

\begin{figure}[!htbp]
\centering
\subfigure[]{
\begin{minipage}[t]{0.5\linewidth}
\centering
\includegraphics[height=6cm,width=8cm]{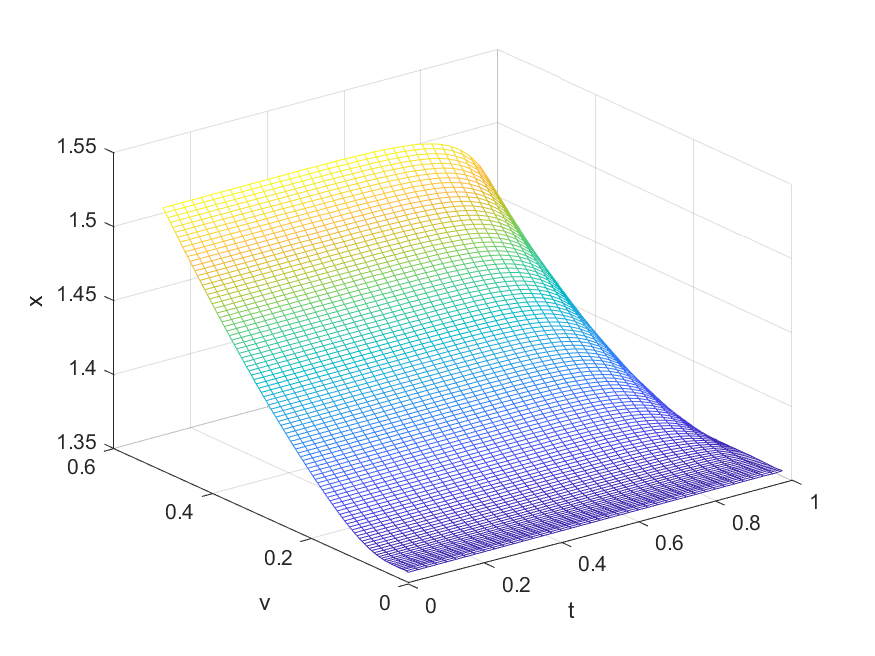}
\end{minipage}%
}%
\subfigure[]{
\begin{minipage}[t]{0.5\linewidth}
\centering
\includegraphics[height=6cm,width=8cm]{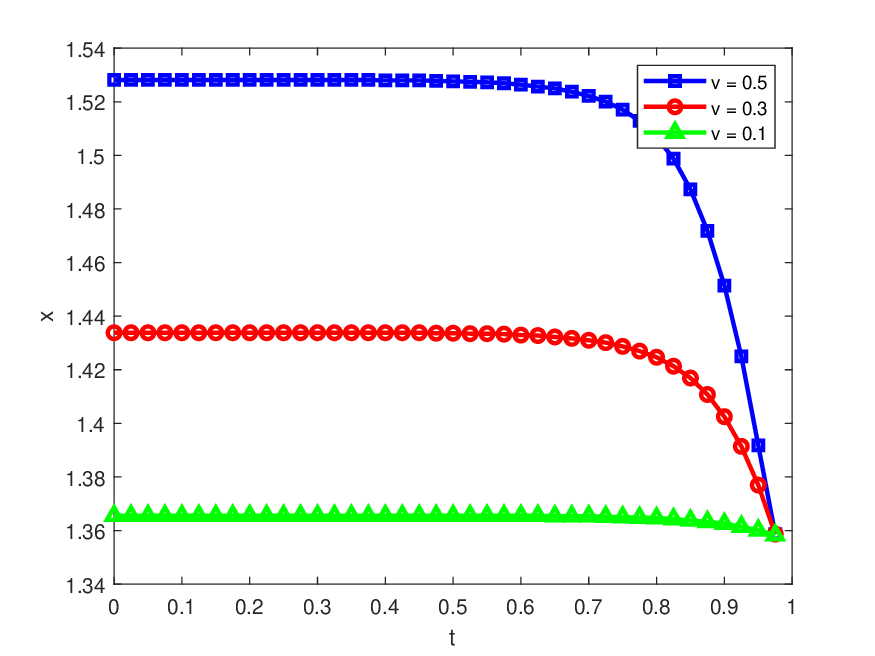}
\end{minipage}%
}

\subfigure[]{
\begin{minipage}[t]{0.5\linewidth}
\centering
\includegraphics[height=6cm,width=8cm]{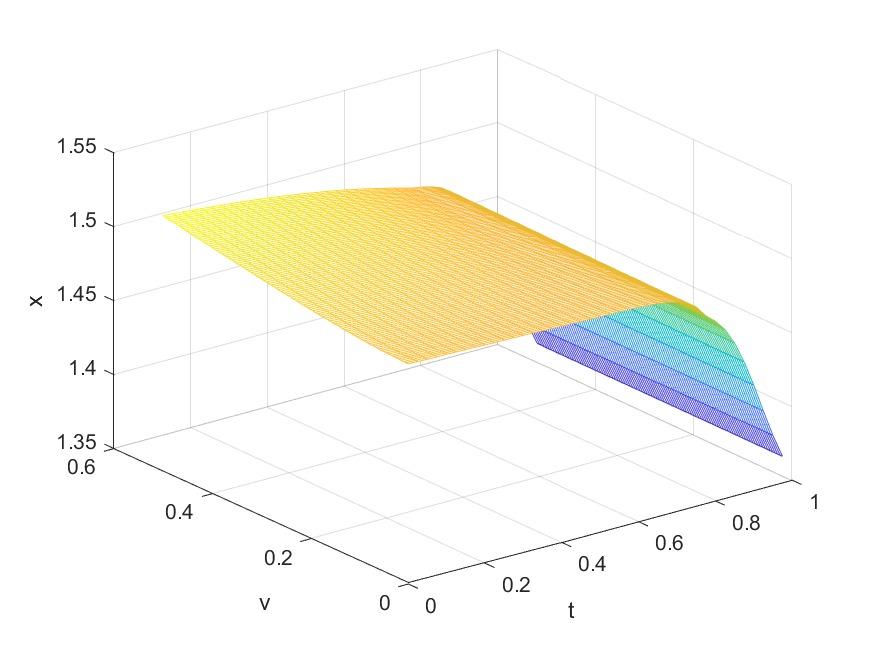}
\end{minipage}%
}%
\subfigure[]{
\begin{minipage}[t]{0.5\linewidth}
\centering
\includegraphics[height=6cm,width=8cm]{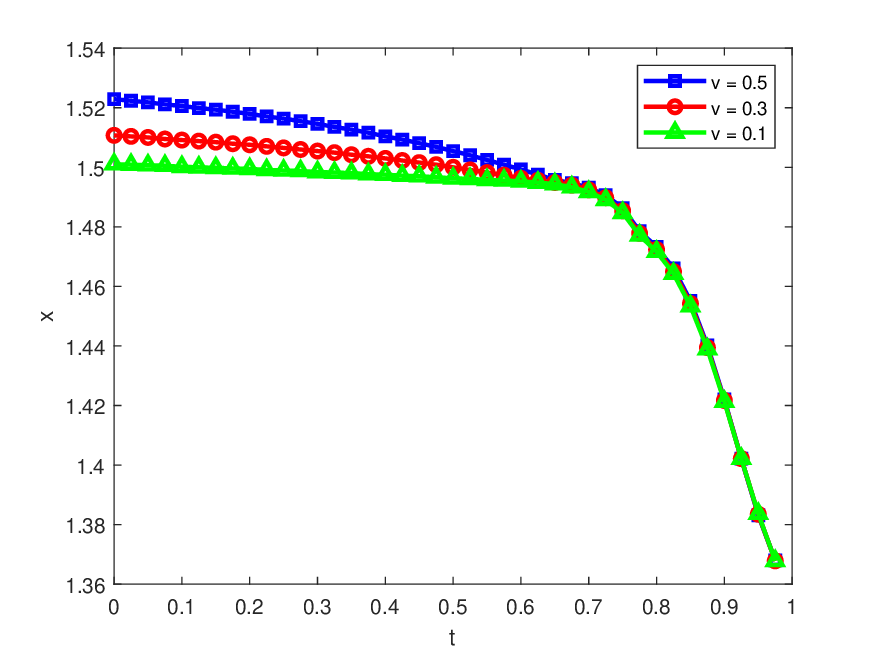}
\end{minipage}%
}

\subfigure[]{
\begin{minipage}[t]{0.5\linewidth}
\centering
\includegraphics[height=6cm,width=8cm]{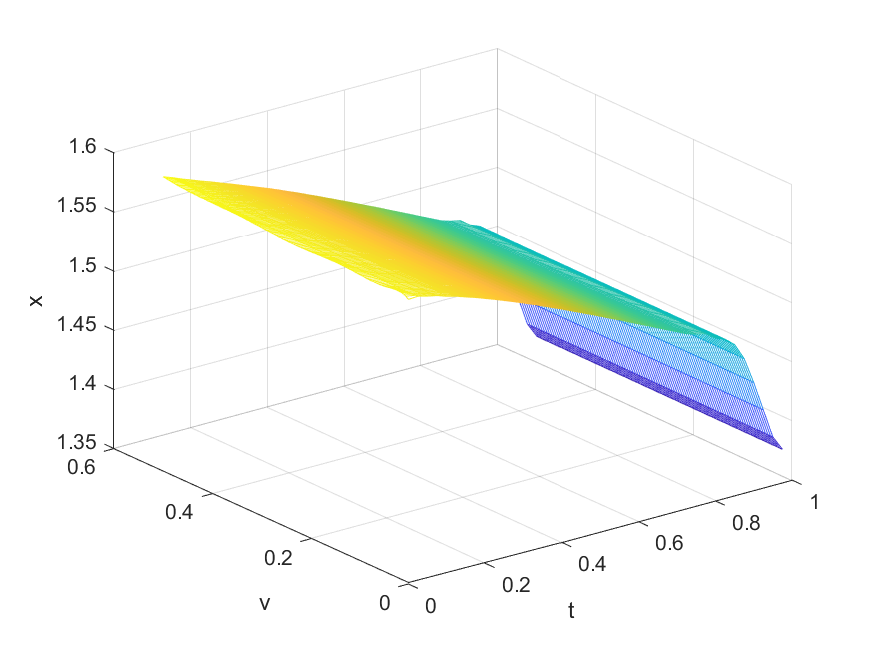}
\end{minipage}%
}%
\subfigure[]{
\begin{minipage}[t]{0.5\linewidth}
\centering
\includegraphics[height=6cm,width=8cm]{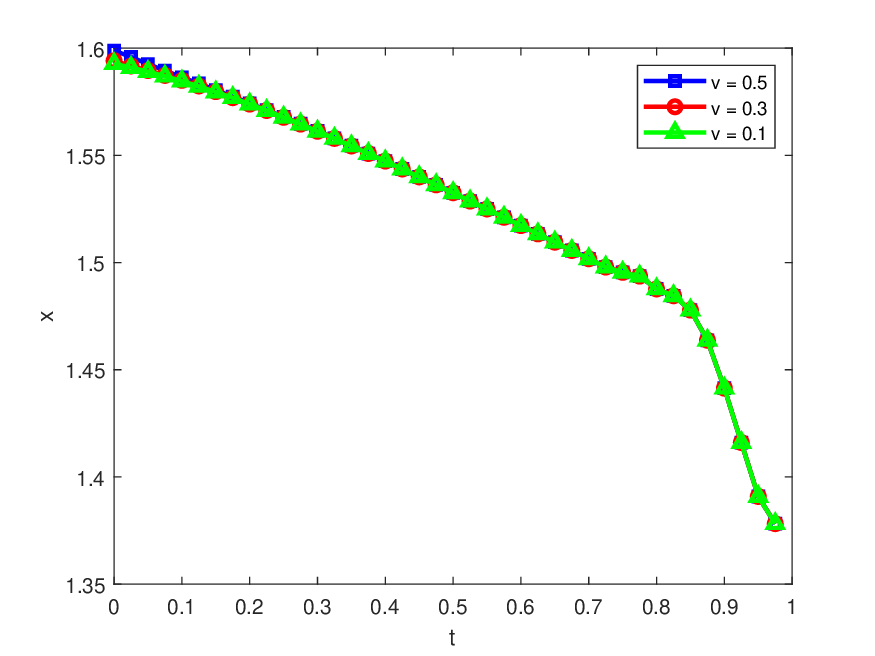}
\end{minipage}%
}
\caption{3D and 2D figures of optimal exercise boundaries (Example \ref{exam:SLVStopping}). Figures (a) and (b) are for Heston models, (c) and (d) for 4/2 models, (e) and (f) for $\alpha$-Hyper models. }
\label{Figure:FreeBoundary}
\end{figure}

In the implementation of the delta family approach, it is taken that
\begin{equation}\label{StoppingSetting}
M=12,\,N=5000,\,x_{\min}=1.2,\,x_{\max}=10,\,v_{\min}=0,\,v_{\max}=1.\notag
\end{equation}
The optimal exercise boundaries are plotted in Figure~\ref{Figure:FreeBoundary}.

\begin{algorithm}[!htbp]
\caption{(Optimal investment with stopping time under SLV models)}\label{alg:SLV-D}
\begin{algorithmic}[1]
\STATE  The algorithm is given by the following recursion for $n=0,1,\ldots,N-1$,
\begin{align}\label{value-SLV-D-algorithm}
\widetilde{V}_{{\bf m}}^{n}&\approx (1-\gamma h)V_{{\bf m}}^{n+1} +\psi_{{\bf m}}^{n+1} \notag\\
&+ h\sum\limits_{k_{0}=0}^{M_0 }\sum\limits_{k_{1}=0}^{M_{1}}\sum\limits_{k_{2}=0}^{M_{2}}
 V_{{\bf k}}^{n+1} \int_{{\Omega}^{3}}\prod\limits_{j=0}^{2}g_{m_{j}}^{(j)}(y_j) {\mathcal L}^{\pi^*_{n+1}}\left(\prod\limits_{i=0}^{2}g_{k_{i}}^{(i)}(y_i)\right)d{\bf y},
\end{align}
where $\pi^*_{n+1}$ is given at \eqref{SLV-optimal-strategy}. Then we have
\begin{equation}\label{approx-value-function-SLV-D}
 \widetilde{V}^{n}({\bf v})\approx\sum\limits_{k_{0}=0}^{M_0 }\sum\limits_{k_{1}=0}^{M_{1}}\sum\limits_{k_{2}=0}^{M_{2}} \widetilde{V}_{{\bf m}}^{n}\prod\limits_{j=0}^{\ell}g_{m_{j}}^{(j)}(\tilde{v_j}(v_j)),\notag
\end{equation}

\begin{equation}
(\psi^n({\bf v}), V^n({\bf v}))=\left\{
 \begin{array}{ll}
\Big(0,\widetilde{V}^n({\bf v})-h\psi^n({\bf v})\Big),&\text{if}~~\widetilde{V}^n({\bf v})-h\psi^{n+1}({\bf v})\geq G,\\
\Big(\frac{G-\widetilde{V}^n({\bf v})}{h}+\psi^{n+1}({\bf v}),G\Big),&\text{otherwise}.
 \end{array}
 \right.\notag
\end{equation}
Then we can calculate
\begin{align}
V^{n}_{{\bf m}}=\int_{\Omega^3}V^n(\hat{\bf v}({\bf y}))\prod\limits_{j=0}^{2}g_{m_{j}}^{(j)}(y_j)d {\bf y},\notag\\
\psi^{n}_{{\bf m}}=\int_{\Omega^3}\psi^n(\hat{\bf v}({\bf y}))\prod\limits_{j=0}^{2}g_{m_{j}}^{(j)}(y_j)d {\bf y}.\notag
\end{align}
The terminal condition
\begin{equation}\label{terminal-condition-SLV-D}
V^{N}_{{\bf m}}=\int_{\Omega^3}U(\hat{v}_0(y_0))\prod\limits_{j=0}^{2}g_{m_{j}}^{(j)}(y_j)d {\bf y}, \qquad\psi^{N}_{{\bf m}}=0.
\end{equation}
The integrals in \eqref{value-SLV-D-algorithm} and \eqref{terminal-condition-SLV-D} can be evaluated by Gauss-Legendre quadrature rules.
\STATE Finally we obtain the approximation of the value functions
\begin{equation}\label{approx-value-function-SLV-D}
 V(t_{n},{\bf v})\approx\sum\limits_{k_{0}=0}^{M_0 }\sum\limits_{k_{1}=0}^{M_{1}}\sum\limits_{k_{2}=0}^{M_{2}} V_{{\bf m}}^{n}\prod\limits_{j=0}^{\ell}g_{m_{j}}^{(j)}(\tilde{v_j}(v_j)),\notag
\end{equation}
and the optimal strategies
\begin{equation}\label{strategy-approx-SLV-D}
\pi^{*}(t_{n},{\bf v})\approx \left\{
 \begin{array}{ll}
\pi^{*}_{n}({\bf v}),&\text{if}~~\psi^n({\bf v})=0,\\
0,&\text{otherwise},
 \end{array}
 \right.\notag
\end{equation}
where $v_0=x$, $v_1=s$, $v_2=v$.
\end{algorithmic}
\end{algorithm}

\section{Conclusions}\label{sec4}
In this paper, we propose a new approach to solve possibly high-dimensional stochastic control (and stopping) problems arising from utility maximization. We provide several algorithms and numerical results to verify the accurateness and efficiency of the proposed approach.
For future research, we shall consider more complex stochastic control problems arising in optimal investment with transaction cost, continuous-time principal-agent problems and mean-field games etc.


\newpage

\begin{thebibliography}{99}
\bibitem[\protect\citeauthoryear{Abi Jaber}{2019}]{Abi2019}
Abi Jaber, E. (2019). Lifting the Heston model. Quantitative Finance, 19, 1995-2013.

\bibitem[\protect\citeauthoryear{Abi Jaber and El Euch}{2019}]{Abi and El2019}
Abi Jaber, E. and El Euch, O. (2019). Multi-factor approximation of rough volatility models. SIAM Journal on Financial Mathematics, 10, 309-349.

\bibitem[\protect\citeauthoryear{Abi Jaber et al.}{2021}]{Abi2021}
Abi Jaber, E., Miller, E. and Pham, H. (2021). Markowitz portfolio selection for multivariate affine and quadratic Volterra models. SIAM Journal on Financial Mathematics, 12, 369-409.

\bibitem[\protect\citeauthoryear{B\"{a}uerle}{2005}]{Bauerle-2005}
B\"{a}uerle, N. (2005). Benchmark and mean-variance problems for insurers. Mathematical Methods of Operations Research, 62, 159-165.

\bibitem[\protect\citeauthoryear{B\"{a}uerle and Desmettre}{2020}]{Bauerle2020}
B\"{a}uerle, N. and Desmettre, S. (2020). Portfolio optimization in fractional and rough Heston models. SIAM Journal on Financial Mathematics, 11, 240-273.


\bibitem[\protect\citeauthoryear{Belomestny et al.}{2010}]{Belo2010}
Belomestny, D., Kolodko, A. and Schoenmakers, J. (2010). Regression methods for stochastic control problems and their convergence analysis. SIAM Journal on Control and Optimization, 48, 3562-3588.

\bibitem[\protect\citeauthoryear{Bi and Cai}{2019}]{Bi and Cai2019}
Bi, J. and Cai, J. (2019). Optimal investment-reinsurance strategies with state dependent risk aversion and VaR constraints in correlated markets. Insurance: Mathematics and Economics, 85, 1-14.

\bibitem[\protect\citeauthoryear{Chen and Shen}{2010}]{Chen2020}
Chen, F. and Shen, J. (2020). Stability and error analysis of operator splitting methods for American options under the Black-Scholes model. Journal of Scientific Computing, 82, 1-17.

\bibitem[\protect\citeauthoryear{Cvitanic and Karatzas}{1992}]{Civitanic1992}
Cvitanic, J. and Karatzas, I. (1992). Convex duality in constrained portfolio optimization. The Annals of Applied Probability, 2, 767-818.


\bibitem[\protect\citeauthoryear{D\'{e}camps and Villeneuve}{2019}]{Decamp2019}
D\'{e}camps, J.P. and Villeneuve, S. (2019). A two-dimensional control problem arising from dynamic contracting theory. Finance and Stochastics, 23, 1-28.


\bibitem[\protect\citeauthoryear{Forsyth and Labahn}{2007}]{Forsyth2007}
Forsyth, P.A. and Labahn, G. (2007). Numerical methods for controlled Hamilton-Jacobi-Bellman PDEs in finance. Journal of Computational Finance, 11, 1-43.


\bibitem[\protect\citeauthoryear{Grasselli}{2017}]{Grasselli2017}
Grasselli, M. (2017). The 4/2 stochastic volatility model: a unified approach for the Heston
and the 3/2 model. Mathematical Finance, 27, 1013-1034.

\bibitem[\protect\citeauthoryear{Gu et al.}{2018}]{Gu2018}
Gu, J.W., Steffensen, M., and Zheng, H. (2018). Optimal dividend strategies of two collaborating businesses in the diffusion approximation model. Mathematics of Operations Research, 43, 377-398.

\bibitem[\protect\citeauthoryear{Guan et al.}{2017}]{Guan2017}
Guan, C., Li, X., Xu, Z.Q., and Yi, F. (2017). A stochastic control problem and related free boundaries in finance. Mathematical Control and Related Fields, 7, 563-584.

\bibitem[\protect\citeauthoryear{Guasoni and Muhle-Karbe}{2013}]{Gua2013}
Guasoni, P. and Muhle-Karbe, J. (2013). Portfolio choice with transaction costs: a user?s guide. In Paris-Princeton Lectures on Mathematical Finance 2013 (pp. 169-201). Springer, Cham.


\bibitem[\protect\citeauthoryear{Han and Wong}{2021}]{Han2021}
Han, B. and Wong, H.Y. (2021). Merton's portfolio problem under Volterra Heston model. Finance Research Letters, 39, 101580.

\bibitem[\protect\citeauthoryear{Han and E}{2016}]{Han2016}
Han, J. and E, W. (2016). Deep learning approximation for stochastic control problems, in Proceedings of NIPS Deep Reinforcement Learning Workshop.

\bibitem[\protect\citeauthoryear{Hur\'{e} et al.}{2021}]
{Pham2021} Hur\'{e}, C., Pham, H., Bachouch, A. and Langren\'{e}, N. (2021). Deep neutal networks algortithms for stochastic control problems on finite horizon: convergence analysis. SIAM Journal on Numerical Analysis, 59, 525-557.


\bibitem[\protect\citeauthoryear{Ikonen}{2004}]{Ikonen2004}
Ikonen, S. and Toivanen, J. (2004). Operator splitting methods for American option pricing. Applied Mathematics Letters, 17,
809-814

\bibitem[\protect\citeauthoryear{Jaber and El Euch}{2019}]{Jaber2019a}
Jaber, E.A. and El Euch, O. (2019). Multi-factor approximation of rough volatility models. SIAM Journal on Financial Mathematics, 10, 309-349.

\bibitem[\protect\citeauthoryear{Jaber}{2019}]{Jaber2019c}
Jaber, E.A. (2019). Lifting the Heston model. Quantitative Finance, 19, 1995-2013.

\bibitem[\protect\citeauthoryear{Kraft}{2005}]{Kraft2005} Kraft, H. (2005). Optimal portfolios and Heston's stochastic volatility
model: an explicit solution for power utility. Quantitative Finance, 5, 303-313.


\bibitem[\protect\citeauthoryear{Lebedev}{1965}]{Lebedev1965}
Lebedev, N.N., (1965). Special Functions and Their Applications. Courier Corporation.

\bibitem[\protect\citeauthoryear{Li and Wong}{2013}]{Li2013}
Li, Y.T., and Wong, R. (2013). Integral and series representations of the Dirac delta function. Communications on Pure and Applied Analysis, 7, 229-247.

\bibitem[\protect\citeauthoryear{Karatzas et al.}{1991}]{Karatzas1991}
Karatzas, I., Lehoczky, J.P., Shreve, S.E. and Xu, G.L. (1991). Martingale and duality methods for utility maximization in an incomplete market. SIAM Journal on Control and optimization, 29, 702-730.

\bibitem[\protect\citeauthoryear{Karatzas and Wang}{2000}]{Ka2000}
Karatzas, I. and Wang, H. (2000). Utility maximization with discretionary stopping. SIAM Journal on Control and Optimization, 39, 306-329.


\bibitem[\protect\citeauthoryear{Ma et al.}{2020}]{Ma2020}
 Ma, J., Li, W. and Zheng, H. (2020). Dual control Monte-Carlo method for tight bounds of value
function under Heston stochastic volatility model. European Journal of Operational Research, 280, 428-440.

\bibitem[\protect\citeauthoryear{Ma et al.}{2022}]{Ma2022}
 Ma, J., Chen, D. and Lu, Z. (2022). Optimal reinsurance-investment with loss aversion under rough Heston model. To appear in Science China Mathematics.

\bibitem[\protect\citeauthoryear{Merton}{1969}]{Merton1969}
Merton, R.C., (1969). Lifetime portfolio selection under uncertainty: The continuous-time case. Review of Economics and Statistics, 51, 247-257.

\bibitem[\protect\citeauthoryear{Pham}{2009}]{Pham2009}
Pham, H. (2009). Continuous-time stochastic control and optimization with financial applications (Vol. 61). Springer Science \& Business Media.

\bibitem[\protect\citeauthoryear{Tahar}{2010}]{Tahar2010}
Tahar, I.B., Soner, H.M. and Touzi, N. (2010). Merton problem with taxes: characterization, computation, and approximation. SIAM Journal on Financial Mathematics, 1, 366-395.


\bibitem[\protect\citeauthoryear{White and Reisinger}{2011}]{White2011}
Witte, J.H. and Reisinger, C. (2011). A penalty method for the numerical solution of Hamilton-Jacobi-Bellman (HJB) equations in finance. SIAM Journal on Numerical Analysis, 49, 213-231.


\bibitem[\protect\citeauthoryear{Yang et al.}{2019}]{Yang2019}
Yang, N., Chen, N. and Wan, X. (2019). A new delta expansion for multivariate diffusions via the Ito-Taylor expansion. Journal of Econometrics, 209, 256-288.

\bibitem[\protect\citeauthoryear{Zhang and Ewald}{2010}]{Zhang-Ewald-2010}
Zhang, A. and Ewald, C.O. (2010). Optimal investment for a pension fund under inflation risk. Mathematical Methods of Operations Research, 71, 353-369.

\end{thebibliography}
\end{document}